\documentclass[reprint, nofootinbib, amsmath,amssymb, aps,floatfix,]{revtex4-2}
\usepackage{gensymb}
\usepackage{textcomp}
\usepackage{lipsum}
\usepackage{graphicx}
\usepackage{dcolumn}
\usepackage{bm}
\usepackage{siunitx}
\DeclareSIUnit\gauss{G}
\DeclareSIUnit\erg{erg}

\usepackage{cancel}
\usepackage{tabularx}
\usepackage{amssymb}
\usepackage{amsmath}
\usepackage{relsize}
\usepackage{listings}
\usepackage{commath}
\usepackage{enumitem}
\usepackage{xfrac}
\usepackage{float}
\usepackage{rotating}
\usepackage{booktabs}
\usepackage{makecell}
\usepackage{mathtools}
\usepackage{caption}
\usepackage{subcaption}
\usepackage{multirow}
\usepackage[table,xcdraw]{xcolor}
\usepackage[version=4]{mhchem}
\usepackage[colorlinks,bookmarks=false,citecolor=red,linkcolor=blue,urlcolor=blue]{hyperref}

\begin{document}

\title{Agent-based Modelling of Quantum Prisoner's Dilemma}
\author{Colin Benjamin}
\email{colin.nano@gmail.com}
\author{Rajdeep Tah}
\email{rajdeep.phys@gmail.com}
\affiliation{School of Physical Sciences, National Institute of Science Education and Research, HBNI, Jatni-752050, India\\
Homi Bhabha National Institute, Training School Complex, Anushakti Nagar, Mumbai-400094, India}

\begin{abstract}
    What happens when an infinite number of players play a quantum game? In this tutorial, we will answer this question by looking at the emergence of cooperation, in the presence of noise, in a one-shot quantum Prisoner's dilemma (QuPD). We will use the numerical Agent-based model (ABM), and compare it with the analytical Nash equilibrium mapping (NEM) technique. To measure cooperation, we consider five indicators, i.e., game magnetization, entanglement susceptibility, correlation, player's payoff average and payoff capacity, respectively. In quantum social dilemmas, entanglement plays a non-trivial role in determining the behaviour of the quantum players (or, \textit{qubits}) in the thermodynamic limit, and for QuPD, we consider the existence of bipartite entanglement between neighbouring quantum players. For the five indicators in question, we observe \textit{first}-order phase transitions at two entanglement values, and these phase transition points depend on the payoffs associated with the QuPD game. We numerically analyze and study the properties of both the \textit{Quantum} and the \textit{Defect} phases of the QuPD via the five indicators. The results of this tutorial demonstrate that both ABM and NEM, in conjunction with the chosen five indicators, provide insightful information on cooperative behaviour in an infinite-player one-shot quantum Prisoner's dilemma.
\end{abstract}

\maketitle

\section{\label{introsec}Introduction}
In the evolutionary context, when we think about examples of social dilemmas (SD), the first thing that comes to our mind is the classical Prisoner's dilemma (or, CPD)\cite{ref1, ref2}. In fact, it is one of the most popular game theoretic models out there that can be used to study a vast array of topics, involving both \textit{one-shot} and \textit{repeated} game settings (see, Refs.~\cite{ref3, ref4, ref5, ref7, ref8}). In CPD, as the word ``\textit{dilemma}" in the name suggests, the Nash equilibrium, i.e., a set of \textit{actions} (or, \textit{strategies}) that lead to an outcome deviating from which one gets worse payoffs, is the \textit{Defect} strategy. This is surprising since there exists a \textit{Pareto optimal} outcome, which for CPD, has better payoffs for both the players and is associated with the \textit{Cooperate} strategy. The \textit{Pareto optimal} strategy and the \textit{Nash equilibrium} strategy are not the same. Quantum game research is seeing a revival of sorts, and recently,  quantum correlation have been explored via Nash equilibrium\cite{lowe} while in \cite{rebecca} quantum advantage is explored via entanglement in two player context.  Till now, most research papers (see, Refs.~\cite{ref3, ref9, ref10, ref11, ref12, ref19}) have been largely restricted to CPD and on understanding how players behave in the \textit{thermodynamic} (or, \textit{infinite} population) \textit{limit} (denoted as \textit{TL}). However, much less focus has been given to the quantum counterpart of CPD, i.e., quantum Prisoner's dilemma (QuPD) in the \textit{TL}. Previously, it was shown, in Refs.~\cite{ref3, ref4, ref5, ref7, ref8, ref15, ref17}, that by quantizing the CPD (see, \textit{Eisert-Wilkens-Lewenstein}, or EWL, protocol in Ref.~\cite{ref8}) and by introducing a unitary \textit{quantum} strategy ($\mathbb{Q}$) in the modified CPD set-up, we can remove the \textit{dilemma} associated with the CPD game. Further research works involving an infinite number of quantum players (see, Refs.~\cite{ref3, ref4, ref7, ref15}) have also shown that the QuPD game can help us understand the emergence of cooperation among the quantum players. However, all these research works have an analytical, rather than a numerical, approach to understanding the question of the emergence of cooperative behaviour. 

In this tutorial, we will numerically study and analyze how cooperative behaviour arises among an infinite number of quantum players playing a \textit{one-shot} QuPD game, and how entanglement ($\gamma$) affects them.  To do so, we will take the help of five different indicators, namely, Game magnetization $(\mu)$, Entanglement susceptibility $(\chi_{\gamma})$, Correlation $(\mathfrak{c}_j)$, Player's payoff average $(\langle \Lambda \rangle)$ and the Payoff capacity $(\wp_C)$, all of them are analogues to the thermodynamic counterparts, i.e., Magnetization, Magnetic susceptibility, Correlation, average Internal energy (or, $\langle \mathbb{E}\rangle$) and the Specific heat capacity at constant volume (or, $\complement_{\mathbb{V}}$), respectively. We will adopt a numerical Agent-based modelling (ABM) technique to study cooperative behaviour among quantum players in the QuPD game, and we will compare our results with the analytical Nash equilibrium mapping (NEM) method. There exist other analytical methods, like Darwinian selection (DS) and Aggregate selection (AS), to analyze $\mu$, $\chi_\gamma$, $\mathfrak{c}_j$, $\langle \Lambda \rangle$ and $\wp_C$, in addition to NEM method. However, in a previous work (see, Ref.~\cite{ref12}), we have shown via a detailed calculation the incorrectness of these analytical methods. Hence, in this work, we will only compare the NEM with the numerical ABM, since both DS and AS are incorrect. Both ABM and NEM are based on the \textit{1D}-Periodic Ising chain (or, IC) with nearest neighbour interactions (see, Refs.~\cite{ref6, ref7, ref10, ref12}). Before moving further, we try to understand what the five aforementioned indicators actually mean. In a symmetric \textit{2-player, 2-strategy} social dilemma game, 2 players (say, $\mathfrak{P}_1$ and $\mathfrak{P}_2$) have 2 different strategies $\$_1$ and $\$_2$ available at their disposal. Thus they have a choice between the two accessible strategies $(\$_1~\text{or}~\$_2)$, which could result in the same or different outcomes (aka, \textit{payoffs}) for each of them. The \textit{four} strategy sets for $\mathfrak{P}_1$ and $\mathfrak{P}_2$, i.e., $(\$|_{\mathfrak{P}_1},~\$|_{\mathfrak{P}_2}) \in \{(\$_1, \$_1),~ (\$_2, \$_1), ~(\$_1, \$_2),~ (\$_2, \$_2)\}$, are each linked to the payoffs $(\mathrm{m, n, p, q})$ via the \textit{symmetric} payoff matrix ($\Lambda$):
\begin{equation}
    \Lambda = \left[\begin{array}{c|c c}
    	 & \$_1 & \$_2\\ 
    	\hline 
    	\$_1 & \mathrm{m,m} & \mathrm{n,p}\\
        \$_2 & \mathrm{p,n} & \mathrm{q,q}
    \end{array}\right]. 
    \label{eq2.0a}
\end{equation}

The \textit{Game magnetization} ($\mu$), analogous to the thermodynamic magnetization, is calculated by subtracting the number of quantum players choosing $\$_2$ strategy from the number of quantum players choosing $\$_1$ strategy. The \textit{Entanglement susceptibility} provides the variation in the rate of change in $\mu$, owing to a change in the entanglement $\gamma$. \textit{Correlation} indicates how closely two quantum players' strategies, at two separate sites, correlate with one another, while the individual player's payoff average, the analogue of $\langle \mathbb{E}\rangle$ for thermodynamic systems, is simply the average payoff that a quantum player receives after playing the game in a \textit{one-shot} environment. Finally, \textit{Payoff capacity}, the analogue of $\complement_{\mathbb{V}}$ for thermodynamic systems, indicates the amount the player's payoff changes for a unit change in \textit{noise}. The \textit{uncertainty} associated with the selection of a strategy by a quantum player is termed \textit{noise}.

We consider all five possible indicators that can be used to numerically study the behaviour of an infinite number of quantum players and construct an agent-based algorithm to study their variation as a function of the entanglement present among the quantum players (or, qubits). We consider the existence of a bipartite entanglement between neighbouring qubits and find two entanglement values at which the five indicators under consideration exhibit first-order phase transitions; the payoffs connected to the quantum Prisoner's dilemma game determine these phase transition points. We systematically examine and investigate the characteristics of the game’s Quantum and Defect phases computationally via the five indicators. The findings of this work show that both the Agent-based modelling method and Nash equilibrium mapping technique, along with all the five indicators, i.e., game magnetization, entanglement susceptibility, correlation, player's payoff average, and payoff capacity, offer informative data on cooperative behaviour in the thermodynamic limit of one-shot quantum Prisoner’s dilemma.

A brief introduction to CPD and QuPD, along with a note on NEM and ABM, is presented in Sec.~\ref{theory}, wherein, we will also understand how QuPD can be mapped to the $1D$-Ising chain (or, IC). In QuPD, for all the five indicators in question, we find that ABM and NEM results exactly match with each other, and they clearly predict the entanglement $\gamma$ range (defined by two critical $\gamma$ values: $\gamma_A$ and $\gamma_B$) in between which, i.e., for $\gamma\in [\gamma_A,\gamma_B]$, \textit{quantum} becomes the dominant strategy. For all indicators we observe an interesting phenomenon of two \textit{first}-order phase transitions, namely, the change of strategies from \textit{Defect} $(\mathcal{D})\rightarrow~\textit{Quantum}~(\mathbb{Q})$ (at entanglement value $\gamma_A$) and $\mathbb{Q}\rightarrow\mathcal{D}$ (at entanglement value $\gamma_B$), regardless of noise in the system. This result is very similar to that observed in Type-\textit{I} superconductors, at a certain critical temperature and in the absence of an external field (see, Refs.~\cite{ref3, ref20}). This also showcases the fact that for QuPD, at finite entanglement $\gamma$ and \textit{zero} noise, we observe a change in the Nash equilibrium condition from \textit{All-$\mathcal{D}$} to \textit{All-$\mathbb{Q}$} and this is marked by a \textit{first}-order phase transition in all the indicators. After analyzing the results obtained for all five indicators via ABM and NEM, we see that all five indicators can identify the phase transition, occurring at two different values of entanglement, i.e., $\gamma_A$ and $\gamma_B$, in QuPD. Remarkably, at finite entanglement, we also see that all five indicators show a phase transition as a function of payoffs too.

The organization of the tutorial is as follows: In Sec.~\ref{theory}, we will discuss both CPD and QuPD, followed by a detailed description of the mathematical framework of NEM and the algorithm of ABM, and how we map our QuPD game to the \textit{1D}-IC. Then, in Sec.~\ref{rsa}, we will discuss and analyze the results obtained for all the five indicators in question, in the \textit{TL}, and finally, we conclude our tutorial by summarizing all the important observations from our work in Sec.~\ref{conc}. 

\section{\label{theory}Theory}
Here, we will discuss both \textit{Classical} and \textit{Quantum} Prisoner's dilemma (PD), followed by a brief introduction to the analytical \textit{Nash equilibrium mapping} (NEM) technique, and finally, we will conclude this section by discussing the algorithm associated to the numerical \textit{Agent-based modelling} (ABM). Both NEM and ABM are based on the exactly solvable $1D$-IC, and instead of dealing with dynamical strategy evolutions, we involve equilibrium statistical mechanics. Hence, we consider a \textit{one-shot} PD game in the infinite-player limit, for both Classical and Quantum cases.  

\subsection{\label{CPD}Classical Prisoner's Dilemma (CPD)}
In CPD \cite{ref1, ref2}, as the name suggests, two \textit{independent} players (say, $\mathfrak{P}_1$ and $\mathfrak{P}_2$), accused of committing a crime, are being interrogated by the law agencies, and they have either option to \textit{Cooperate} ($\mathfrak{C}$) with each other or \textit{Defect} ($\mathfrak{D}$). If both players opt for $\mathfrak{C}$-strategy, then they are rewarded with a payoff $\mathbb{R}$, whereas, if both choose $\mathfrak{D}$-strategy, then they get the punishment payoff $\mathbb{P}$. However, if both $\mathfrak{P}_1$ and $\mathfrak{P}_2$ choose opposite strategies, then the one choosing $\mathfrak{C}$-strategy gets the \textit{sucker's} payoff $\mathbb{S}$, and the one choosing $\mathfrak{D}$-strategy gets the temptation payoff $\mathbb{T}$, respectively. The payoffs have to fulfil the criteria: $\mathbb{T}>\mathbb{R}>\mathbb{P}>\mathbb{S}$. Hence, the CPD payoff matrix ($\Tilde{\Xi}$) is,
\begin{equation}
    \Tilde{\Xi} = \left[\begin{array}{c|c c}
    	 & \mathfrak{C} & \mathfrak{D}\\ 
    	\hline 
    	\mathfrak{C} & \mathbb{R, R} & \mathbb{S, T}\\
        \mathfrak{D} & \mathbb{T, S} & \mathbb{P, P}
    \end{array}\right]. 
    \label{eq2.1}
\end{equation}
From $\Tilde{\Xi}$ in Eq.~(\ref{eq2.1}), one would think that the two rational players, who are always looking for payoff maximization, would choose the Pareto optimal $\mathfrak{C}$-strategy since it is a \textit{win-win} situation for both. However, owing to independence in strategy selection, both $\mathfrak{P}_1$ and $\mathfrak{P}_2$ choose the $\mathfrak{D}$-strategy (thus the name ``\textit{dilemma}") to ensure that none of them receives the minimum payoff, i.e., the sucker's payoff $\mathbb{S}$, due to a unilateral change in the opponent's strategy. Hence, the Nash equilibrium in CPD is $\mathfrak{D}$. For our case, we rewrite the CPD payoffs $\{\mathbb{R, S, T, P}\}$ in terms of a new set of payoffs: Cooperation bonus ($\mathbb{B}$) and Cost ($\mathbb{C}$), where we redefine $\mathbb{R} = \mathbb{B-C}$ (i.e., cooperation bonus with the cost subtracted out), $\mathbb{T = B}$ (or, the entire Cooperation bonus), $\mathbb{S = -C}$ (i.e., bearing the cost without any bonus) and $\mathbb{P}=0$, respectively. Here, $\mathbb{B}\geq \mathbb{C}\geq 0$, with $\mathbb{B}\geq (\mathbb{B-C})\geq 0 \geq -\mathbb{C}$ (i.e., $\mathbb{T}>\mathbb{R}>\mathbb{P}>\mathbb{S}$ criteria is satisfied in this case also), and we will be using the same $\{\mathbb{B, C}\}$ as our payoffs when we deal with the quantum Prisoner's dilemma in the next section.   

\subsection{\label{QPD}Quantum Prisoner's Dilemma (QuPD)}
The topic of extending the framework of CPD to the quantum regime is discussed elaborately in Refs.~\cite{ref3, ref4, ref5, ref8}. Still, we will discuss them in this tutorial for the reader's convenience. The players in CPD, say $\mathfrak{P}_1$ and $\mathfrak{P}_2$, are now treated as \textit{qubits}, i.e., quantum players, in QuPD, and the state of the qubits represent the strategies adopted by the quantum players. Analogous to CPD, in QuPD, the \textit{Cooperate}-strategy is denoted by $|\mathfrak{C}\rangle$ and the \textit{Defect}-strategy is denoted by $|\mathfrak{D}\rangle$, respectively. Note that even though the quantum players are denoted by \textit{qubits}, the strategies available to the quantum players are still classical in nature. Here, 
\begin{equation}
    |\mathfrak{C}\rangle = [1~~0]^{\perp}, ~\text{and}~|\mathfrak{D}\rangle = [0~~1]^{\perp},
    \label{eq2.2}
\end{equation}
where, $\perp$ denotes \textit{transpose} of the given row matrices. The strategy to be adopted by every quantum player is given by acting the unitary operator \cite{ref8},
\begin{equation}
    \hat{\mathfrak{U}}(\varphi,\vartheta) = \begin{bmatrix}
        e^{i\varphi}\cos{(\frac{\vartheta}{2})} & \sin{(\frac{\vartheta}{2})}\\
        -\sin{(\frac{\vartheta}{2})} & e^{-i\varphi}\cos{(\frac{\vartheta}{2})}
    \end{bmatrix},
    \label{eq2.3}
\end{equation}
on the initial state of the quantum players. The operator $\hat{\mathfrak{U}}(\varphi,\vartheta)$ $\forall~\vartheta\in [0,\pi],~\varphi \in [0,\frac{\pi}{2}]$, acts independently on the individual Hilbert spaces of both quantum players. Here, $\hat{\mathfrak{U}}(\varphi = 0,\vartheta=0) = \hat{\mathbb{I}}_{2\times 2}$ (i.e., \textit{Identity} operator) represents the \textit{cooperation} strategy, and 
\begin{equation*}
    \hat{\mathfrak{U}}(\varphi = 0,\vartheta=\pi) = \begin{bmatrix}
        0 & 1\\
        -1 & 0
    \end{bmatrix},
\end{equation*}
represents the \textit{defection} strategy \cite{ref3, ref8}. Since we are dealing with QuPD, the concept of \textit{entanglement} comes into the picture, and before any strategy-modification operation, the two \textit{distinct}, \textit{individual} qubits of the two quantum players are entangled via the \textit{unitary} entanglement operator \cite{ref8},
\begin{gather}
    \hat{\Gamma} (\gamma) = \bigg(\cos{\frac{\gamma}{2}}\bigg) \hat{\mathbb{I}}\otimes\hat{\mathbb{I}} - \bigg(i\sin{\frac{\gamma}{2}}\bigg) \hat{\mathfrak{S}}_Y\otimes\hat{\mathfrak{S}}_Y,
    \label{eq2.4}
\end{gather}
where, $\hat{\mathbb{I}}$ is the $2\times 2$ Identity operator, $\hat{\mathfrak{S}}_Y = \begin{bmatrix}
        0 & -i\\
        i & 0
    \end{bmatrix}$ is the \textit{Pauli-Y} operator, and $\gamma$ denotes the \textit{entanglement parameter} in QuPD, i.e., for $\gamma = \frac{\pi}{2}$, we have \textit{maximal} entanglement and for $\gamma=0$, we have \textit{minimal} entanglement. Both the quantum players are well aware of the entanglement operator $\hat{\Gamma} (\gamma)$. If both the \textit{quantum players/qubits} (say, $\mathfrak{P}_1$ and $\mathfrak{P}_2$) choose $|\mathfrak{C}\rangle$ as their initial states and $\hat{\Gamma} (\gamma)$ acting on them gives the entangled state (from, Eq.~(\ref{eq2.4})) \cite{ref8},
\begin{equation}
    |\alpha\rangle =\hat{\Gamma} (\gamma) |\mathfrak{CC}\rangle  = \cos{\frac{\gamma}{2}} |\mathfrak{CC}\rangle + i\sin{\frac{\gamma}{2}} |\mathfrak{DD}\rangle.
    \label{eq2.5}
\end{equation}
Subsequently, both quantum players apply $\hat{\mathfrak{U}}(\varphi,\vartheta)$ on their respective qubits, giving us the intermediate state \cite{ref8},
\begin{equation}
    |\beta\rangle = \hat{\mathfrak{U}}(\varphi_{\mathfrak{P}_1},\vartheta_{\mathfrak{P}_1})\otimes \hat{\mathfrak{U}}(\varphi_{\mathfrak{P}_2},\vartheta_{\mathfrak{P}_2}) \hat{\Gamma} (\gamma) |\mathfrak{CC}\rangle.
    \label{eq2.6}
\end{equation}
Finally, before any measurement is made, the disentangling operator $\hat{\Gamma}^\dagger (\gamma)$ is acted on $|\beta\rangle$, given in Eq.~(\ref{eq2.6}), and the final non-entangled state is given as \cite{ref8},
\begin{equation}
    |\mathbb{F}\rangle = \hat{\Gamma}^\dagger (\gamma) \hat{\mathfrak{U}}(\varphi_{\mathfrak{P}_1},\vartheta_{\mathfrak{P}_1})\otimes \hat{\mathfrak{U}}(\varphi_{\mathfrak{P}_2},\vartheta_{\mathfrak{P}_2}) \hat{\Gamma} (\gamma) |\mathfrak{CC}\rangle.
    \label{eq2.7}
\end{equation}
Similar to CPD, we can determine the game payoffs for QuPD by projecting the final non-entangled state $|\mathbb{F}\rangle$ onto the basis vectors that entirely span the combined Hilbert space of the two-qubits, i.e., $|\mathfrak{CC}\rangle$, $|\mathfrak{DD}\rangle$, $|\mathfrak{CD}\rangle$ and $|\mathfrak{DC}\rangle$, respectively. These values, coupled with the CPD payoffs (see, Eq.~(\ref{eq2.1}) and new payoffs $\mathbb{B, C}$), give us the QuPD payoffs ($\Lambda$) for both quantum players: $\mathfrak{P}_1$ and $\mathfrak{P}_2$, as,
\begin{gather}
    \Lambda_{\mathfrak{P}_1} = (\mathbb{B-C}) \sigma_{\mathfrak{CC}} - \mathbb{C} \sigma_{\mathfrak{CD}} + \mathbb{B} \sigma_{\mathfrak{DC}} + \cancelto{0}{\mathbb{P}}\sigma_{\mathfrak{DD}} \label{eq2.8}\\
    \Lambda_{\mathfrak{P}_2} = (\mathbb{B-C}) \sigma_{\mathfrak{CC}} - \mathbb{C} \sigma_{\mathfrak{DC}} + \mathbb{B} \sigma_{\mathfrak{CD}} + \cancelto{0}{\mathbb{P}}\sigma_{\mathfrak{DD}}
    \label{eq2.9}
\end{gather}
where, $\sigma_{\mathfrak{CC}} = |\langle \mathbb{F}|\mathfrak{CC} \rangle|^2$, $\sigma_{\mathfrak{CD}} = |\langle \mathbb{F}|\mathfrak{CD} \rangle|^2$, $\sigma_{\mathfrak{DC}} = |\langle \mathbb{F}|\mathfrak{DC} \rangle|^2$, and $\sigma_{\mathfrak{DD}} = |\langle \mathbb{F}|\mathfrak{DD} \rangle|^2$, respectively. By introducing a \textit{Quantum} strategy operator \cite{ref3}, $\hat{\mathbb{Q}} = i \hat{\mathfrak{S}}_Z = \hat{\mathfrak{U}}(\varphi = \frac{\pi}{2},\vartheta = 0)$, where, $\hat{\mathfrak{S}}_Z = \begin{bmatrix}
        1 & 0\\
        0 & -1
    \end{bmatrix}$ is the \textit{Pauli-Z} operator, we have the quantum players opting for a different strategy than the available classical strategies $\mathfrak{C}$ and $\mathfrak{D}$. Initially, superposition states like $\frac{1}{\sqrt{2}}(|\mathfrak{C}\rangle + |\mathfrak{D}\rangle)$ used to remain invariant under the action of classical strategies. However, upon the action of $\hat{\mathbb{Q}}$ on $\frac{1}{\sqrt{2}}(|\mathfrak{C}\rangle + |\mathfrak{D}\rangle)$, we get an orthogonal state $\frac{1}{\sqrt{2}}e^{i\pi/2}(|\mathfrak{C}\rangle - |\mathfrak{D}\rangle)$, where $e^{i\pi/2}$ is a global phase factor. In QPD, we involve both quantum and classical strategies, and the payoff matrix is:
\begin{equation}
    \Tilde{\Lambda} = \left[\begin{array}{c|c c c} 
    	 & {\mathfrak{C}} & {\mathfrak{D}} & {\mathbb{Q}}\\ 
    	\hline 
    	{\mathfrak{C}} & \mathbb{B-C,B-C} & \mathbb{-C,B} & \mathfrak{L_1, L_1}\\
        {\mathfrak{D}} & \mathbb{B,-C} & {0,0} & \mathfrak{L_3, L_2}\\
        {\mathbb{Q}} & \mathfrak{L_1, L_1} & \mathfrak{L_2, L_3} & \mathbb{B-C,B-C}
    \end{array}\right].
    \label{eq2.10}
\end{equation}
\begin{figure*}[!ht]
    \centering
    \includegraphics[width = 0.95\textwidth]{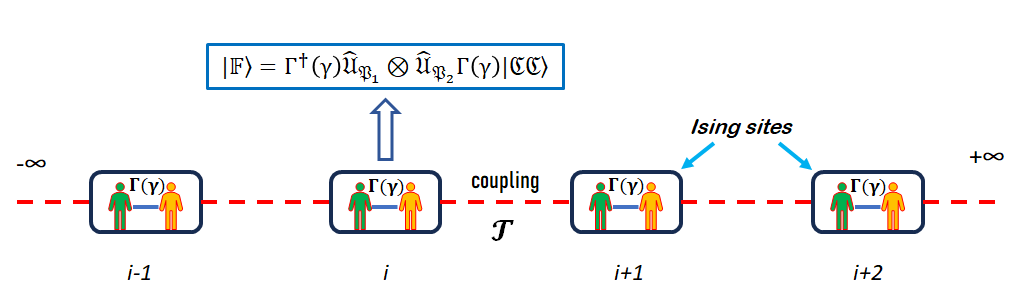}
    \caption{\centering{\textbf{NEM/ABM}: Visualization of QuPD in thermodynamic (or, \textit{infinite-player}) limit. The QuPD game is mapped to a $1D$-Ising chain, where at each site, two entangled quantum players reside, and they play the \textit{two}-strategy QuPD. The quantum players also interact with their nearest neighbours via a classical coupling $\mathcal{T}$, in the presence of an external uniform field $\mathcal{F}$ and \textit{noise} $\beta$.}}
    \label{fig:0}
\end{figure*}
where, $\mathfrak{L_1} = (\mathbb{B-C})\cos^2{\gamma}$, $\mathfrak{L_2} = -\mathbb{C}\cos^2{\gamma} + \mathbb{B}\sin^2{\gamma}$, and $\mathfrak{L_3} = \mathbb{B}\cos^2{\gamma} - \mathbb{C}\sin^2{\gamma}$. Hence, QPD is a \textit{three}-strategy game, but it is difficult to map the \textit{three}-strategy QuPD to an exactly solvable \textit{spin-1} IC \cite{ref3}. However, previous works on QuPD (see, Refs.~\cite{ref3, ref5, ref8}) have shown how $\mathbb{Q}$ fare against the classical strategies $\mathfrak{C}$ and $\mathfrak{D}$, indicating that it will be a wiser choice to divide our \textit{three}-strategy QuPD into two \textit{two}-strategy QuPD problems (so that we can map them individually to a \textit{spin-1/2} IC), and we can compare the ($\mathbb{Q}$ vs $\mathfrak{D}$) as well as the ($\mathbb{Q}$ vs $\mathfrak{C}$) cases. As visualized in Fig.~\ref{fig:0}, for each case of the \textit{two}-strategy QuPD, we have two entangled quantum players (playing the \textit{two}-strategy QuPD) at every site of the infinitely long $1D$-IC (i.e., the \textit{thermodynamic} limit), and each site is coupled to its nearest neighbouring sites via the coupling constant $\mathcal{T}$. All the sites, each consisting of two entangled quantum players, are subjected to a uniform external field $\mathcal{F}$ (analogous to the external magnetic field Ising factor). From Eq.~(\ref{eq2.10}), one can notice from the given payoff matrix $\Lambda$ that for ($\mathbb{Q}$ vs $\mathfrak{C}$) case, both $(\mathbb{Q, Q})$ and $(\mathfrak{C, C})$ strategy pairs gives the same payoff to both the quantum players, and we observe no payoff variations when switching from classical to quantum strategies. Owing to this, we only consider the ($\mathbb{Q}$ vs $\mathfrak{D}$) case for our further work.

For ($\mathbb{Q}$ vs $\mathfrak{D}$) case, we have the $2\times 2$-reduced payoff matrix for a single quantum player (say, the \textit{row} player) as,
\begin{equation}
    \Lambda = \left[\begin{array}{c|c c} 
    	 & {\mathbb{Q}} & {\mathfrak{D}}\\ 
    	\hline 
        {\mathbb{Q}} & (\mathbb{B-C}) & ( \mathbb{B}\sin^2{\gamma}-\mathbb{C}\cos^2{\gamma})\\
        {\mathfrak{D}} & (\mathbb{B}\cos^2{\gamma} - \mathbb{C}\sin^2{\gamma}) & 0
    \end{array}\right].
    \label{eq2.11}
\end{equation}
This $\Lambda$, in Eq.~(\ref{eq2.11}), will be utilised in our further calculations in the thermodynamic limit. In the next section, we will discuss the analytical NEM very briefly, and we will compare the results from NEM with the results of numerical ABM.

\subsubsection{\label{NEM}Nash equilibrium mapping (NEM)}
In one of our previous works (see, Ref.~\cite{ref12}), we have discussed the mathematical framework of NEM in great detail. To summarize for the readers, in NEM, we analytically map a SD to a \textit{spin-1/2} infinitely long $1D$-IC (i.e., the \textit{TL}) (see, Refs.~\cite{ref6, ref10, ref12}). Initially, we consider a $2$-strategy; $2$-player SD while mapping it to a $2$-site IC. The 2 strategies (say, $\$_1$ and $\$_2$) have a \textit{one-to-one} mapping to the \textit{2}-spin (say, $\pm1$) $1D$-IC and we have the $2$-site $(\text{say,}~A~\text{and}~B)$ $1D$-IC Hamiltonian as \cite{ref12},
\begin{equation}
    H = -\mathcal{T}(\mathfrak{\sigma}_A\mathfrak{\sigma}_B + \mathfrak{\sigma}_B\mathfrak{\sigma}_A) - \mathcal{F}(\mathfrak{\sigma}_A + \mathfrak{\sigma}_B) = \Delta_A + \Delta_B,
    \label{eq2.12}
\end{equation}
where, $\mathcal{T}$ is the coupling constant, $\mathcal{F}$ is the external field, $\mathfrak{\sigma}_i ~\forall~i\in \{A,B\},$ is the spin (either $+1$ or $-1$) at the $i^{th}$ site, and $\Delta_i$ denotes the energy of the $i^{th}$ site. Each of the individual site's energy is given as,
\begin{equation}
    \Delta_A = -\mathcal{T}\mathfrak{\sigma}_A\mathfrak{\sigma}_B - \mathcal{F}\mathfrak{\sigma}_A,~\text{and}~\Delta_B = -\mathcal{T}\mathfrak{\sigma}_B\mathfrak{\sigma}_A - \mathcal{F}\mathfrak{\sigma}_B.
    \label{eq2.13}
\end{equation}
Here, the total two-spin IC energy: $\Delta= \Delta_A +\Delta_B$. In social dilemmas, the maximization of the player's feasible payoffs corresponds to finding the Nash equilibrium of the game. However, when we consider a $1D$-IC, we look to lessen the $\Delta$ in order to reach the energy equilibrium condition. Hence, in order to establish a link between the IC's energy equilibrium configuration and the Nash equilibrium of a SD, we equate the SD payoff matrix to the negative of the energy matrix \cite{ref6}. Each element of $\Delta$ (i.e., $\Delta_i$) corresponds to a particular pair of spin values $(\sigma_A,\sigma_B)$, i.e., $(\sigma_A,\sigma_B)\in \{(\pm 1, \pm 1)\}$, at each $1D$-IC site since game payoff maximization indicates negative energy minimization. Thus,
\begin{equation}
\begin{footnotesize}
    -\Delta = 
    \left[\begin{array}{c|c c} 
    	 & \mathfrak{\sigma}_B = +1 & \mathfrak{\sigma}_B = -1\\ 
    	\hline 
    	\mathfrak{\sigma}_A = +1 & (\mathcal{T}+\mathcal{F}), (\mathcal{T}+\mathcal{F}) & (\mathcal{F}-\mathcal{T}), -(\mathcal{F}+\mathcal{T})\\
        \mathfrak{\sigma}_A = -1 & -(\mathcal{F}+\mathcal{T}), (\mathcal{F}-\mathcal{T}) & (\mathcal{T}-\mathcal{F}), (\mathcal{T}-\mathcal{F})
    \end{array}\right].
    \label{eq2.14}
\end{footnotesize} 
\end{equation}
For a $2$-player; $2$-strategy (say, $\$_1$ and $\$_2$) \textit{symmetric} SD game, we have the SD payoff matrix $\Lambda'$ as,
\begin{equation}
    \Lambda' = \left[\begin{array}{c|c c} 
    	 & \$_1 & \$_2\\ 
    	\hline 
    	\$_1 & \mathrm{m, m} & \mathrm{n, p}\\
        \$_2 & \mathrm{p, n} & \mathrm{q, q}
    \end{array}\right]. 
    \label{eq2.15}
\end{equation}
where, $(\mathrm{m,n,p, q})$ are defined as the SD payoffs. Using a set of linear transformations on $\Lambda'$ in Eq.~(\ref{eq2.15}), to establish a \textit{one-to-one} correspondence between the payoffs in Eq.~(\ref{eq2.15}) and the energy matrix $\Delta$ of the two-spin Ising chain in Eq.~(\ref{eq2.14}), that preserves the Nash equilibrium (see, Ref.~\cite{ref6} and \textbf{Appendix} of Ref.~\cite{ref5a} for detailed calculations),
\begin{equation}
    \mathrm{m}\rightarrow \frac{\mathrm{m}-\mathrm{p}}{2},~\mathrm{n} \rightarrow \frac{\mathrm{n}-\mathrm{q}}{2},~\mathrm{p}\rightarrow \frac{\mathrm{p}-\mathrm{m}}{2},~\mathrm{q}\rightarrow \frac{\mathrm{q}-\mathrm{n}}{2},
    \label{eq2.16}
\end{equation}
we equate $\Lambda'$ (see, Eq.~(\ref{eq2.15})) to $-\Delta$ (see, Eq.~(\ref{eq2.14})), to rewrite the Ising parameters $(\mathcal{T},\mathcal{F})$ in terms of the SD payoffs as \cite{ref6, ref10, ref12},
\begin{equation}
    \mathcal{F} = \frac{\mathrm{(m-p)+(n-q)}}{4},~ \text{and}~\mathcal{T} = \frac{\mathrm{(m-p)-(n-q)}}{4}.
    \label{eq2.17}
\end{equation}
In $1D$-IC, the parameter $\beta$ is defined as proportional to the temperature ($T$) inverse, or, $\beta = \frac{1}{k_B T}$, where $k_B$ is the \textit{Boltzmann constant}, and in game theoretic models, temperature implies \textit{uncertainty} in quantum player's strategy selection, and this is termed as \textit{noise}. Therefore, $T\rightarrow 0$ (or, $\beta \rightarrow \infty$) implies \textit{zero noise} (denoted by Z-N), i.e., no change in the quantum players' strategies, whereas, $T\rightarrow \infty$ (or, $\beta \rightarrow 0$) implies \textit{infinite noise} (denoted by I-N), i.e., complete randomness in the quantum player's strategy selection. We can also interpret $\beta$ as the \textit{selection intensity} (see, Ref.~\cite{ref13}) where, for $\beta\ll 1$, we observe strategies being selected at random, whereas, for $\beta\gg 1$, we have a vanishing randomness in strategy selection. 

For the given $H$ (see, Eq.~(\ref{eq2.12})) and Eq.~(\ref{eq2.17}), the partition function $\Upsilon^{NEM}$ can be written in terms of the SD parameters $(\mathrm{m, n, p, q})$ as \cite{ref12},
\begin{gather}
    \Upsilon^{Ising} = \sum_{\{\sigma_i\}} e^{-\beta H} = e^{2\beta (\mathcal{T}+\mathcal{F})} + e^{2\beta (\mathcal{T}-\mathcal{F})} + 2e^{-2\beta\mathcal{T}},\nonumber\\
    \text{or,}~\Upsilon^{NEM} = e^{\beta\mathrm{(m-p)}} + e^{-\beta \mathrm{(n+q)}} + 2e^{\frac{\beta}{2} \mathrm{(n+p-m-q)}},
    \label{eq2.18}
\end{gather}
where, $\beta = \frac{1}{k_B T}$. For ($\mathbb{Q}$ vs $\mathfrak{D}$) case of QuPD, from Eq.~(\ref{eq2.11}), we have: 
\begin{gather}
    \mathrm{m} = (\mathbb{B-C});~~\mathrm{n} = ( \mathbb{B}\sin^2{\gamma}-\mathbb{C}\cos^2{\gamma});\nonumber\\
    \mathrm{p} = (\mathbb{B}\cos^2{\gamma} - \mathbb{C}\sin^2{\gamma});~~\text{and}~~\mathrm{q} = 0.
    \label{eq2.19}
\end{gather}
Hence, from Eqs.~(\ref{eq2.17}, \ref{eq2.19}), we have $(\mathcal{T}, \mathcal{F})$ in terms of the SD payoffs $(\mathbb{B}, \mathbb{C})$ and entanglement $\gamma$ as,
\begin{gather}
    \mathcal{T} = 0,~~\text{and}~~\mathcal{F} = \dfrac{(\mathbb{B}\sin^2{\gamma}-\mathbb{C}\cos^2{\gamma})}{2}
    \label{eq2.20}
\end{gather}
The readers can refer to our previous work, involving NEM; ABM and other analytical methods, in Ref.~\cite{ref12}, where the detailed calculations on deriving the analytical expressions for the given five indicators of cooperative behaviour: game magnetization $(\mu^{NEM})$, entanglement susceptibility $(\chi_\gamma^{NEM})$, correlation $(\mathfrak{c}_j^{NEM})$, player's payoff average $(\langle \Lambda\rangle^{NEM})$ and payoff capacity $(\wp_C^{NEM})$, with regards to two different classical SD (Hawk-Dove game and Public goods game) are shown. The same technique, as in Ref.~\cite{ref12}, is followed here. So, we have the analytical expressions for the five different indicators for QuPD, in the thermodynamic (or, \textit{infinite quantum players}) limit, as,\\
\textbf{1. Game magnetization:} Using the $\Lambda$ in Eq.~(\ref{eq2.11}) and $\Upsilon^{NEM}$ (see, Eqs.~(\ref{eq2.18}, \ref{eq2.20})), we have the QuPD average game magnetization $\mu^{NEM}$ in the \textit{TL} as, 
\begin{gather}
    \mu^{NEM} = \dfrac{1}{\beta}\dfrac{\partial}{\partial \mathcal{F}}\ln{\Upsilon^{NEM}} = \dfrac{\sinh{\beta\mathcal{F}}}{\sqrt{e^{-4\beta \mathcal{T}} + \sinh^2{\beta\mathcal{F}}}},\nonumber\\
    \text{or,}~~ \mu^{NEM}=\dfrac{\sinh{\bigg[\dfrac{\beta}{2} (\mathbb{B}\sin^2{\gamma}-\mathbb{C}\cos^2{\gamma})\bigg]}}{\mathfrak{W}},
    \label{eq2.21}    
\end{gather}
where, $\mathfrak{W} = \sqrt{\sinh^2{[\frac{\beta}{2} (\mathbb{B}\sin^2{\gamma}-\mathbb{C}\cos^2{\gamma})]} + 1}$, respectively. Here, $\mathbb{B}$ is the Cooperation bonus, $\mathbb{C}$ is the cost, and $\gamma$ denotes the entanglement associated with the QuPD game.\\ 
\textbf{2. Entanglement susceptibility:} To derive the analytical expression for the QuPD Entanglement susceptibility $\chi_\gamma^{NEM}$, we partially differentiate $\mu^{NEM}$ with the entanglement parameter $\gamma$ and normalize it by $\beta$, i.e.,
\begin{gather}
    \chi_\gamma^{NEM} = \dfrac{1}{\beta}\dfrac{\partial}{\partial \gamma}\mu^{NEM}~\text{, for}~\gamma\in [0, \pi ],\nonumber\\
    \text{or,}~~\chi_\gamma^{NEM} = \dfrac{(\mathbb{B+C})}{2}\cdot \dfrac{\sin{(2\gamma)} \cosh{(\beta \mathcal{N})}}{(1 + \sinh^2{\beta \mathcal{N}})^{\frac{3}{2}}},
    \label{eq2.22}    
\end{gather}
where, $\mathcal{N} = \frac{1}{2}(\mathbb{B}\sin^2{\gamma}-\mathbb{C}\cos^2{\gamma})$, respectively.\\
\textbf{3. Correlation:} In our previous work in Ref.~\cite{ref12}, we derived the analytical NEM expression for the correlation $\mathfrak{c}_j^{NEM} (= \langle \sigma_i \sigma_{i+j}\rangle)$ between two sites separated by a distance $j$, for classical social dilemmas (CSD), as,
\begin{equation}
    \mathfrak{c}_j^{NEM}\bigg|_{CSD} =  \cos^2 \varphi + \bigg(\dfrac{\Omega_{-}}{\Omega_{+}} \bigg)^j \sin^2 \varphi,
    \label{eq2.24}
\end{equation}
with $\cos^2 \varphi= \frac{\sinh^2{\beta \mathcal{F}}}{\mathfrak{Y}},~\Omega_{\pm} = e^{\beta \mathcal{T}} [\text{cosh}(\beta \mathcal{F}) \pm \sqrt{\mathfrak{Y}}]$, where $\mathfrak{Y} = \sinh^2{\beta \mathcal{F}} + e^{-4\beta \mathcal{T}}$. In QuPD, the calculation for correlation is very similar, where we just replace the $(\mathcal{T}, \mathcal{F})$ given in Eq.~(\ref{eq2.24}) with the values given in Eq.~(\ref{eq2.20}) to have the expression for the QuPD correlation $\mathfrak{c}_j^{NEM} = \langle \sigma_i \sigma_{i+j}\rangle$ as,
\begin{equation}
    \mathfrak{c}_j^{NEM} = \cos^2 \Phi + \bigg(\dfrac{\omega_{-}}{\omega_{+}} \bigg)^j \sin^2 \Phi,
    \label{eq2.25}    
\end{equation}
where, $j$ is the \textit{distance} between the two sites and
\begin{gather}
    \cos^2 \Phi = \dfrac{\sinh^2{\beta\mathcal{N}}}{\mathfrak{W}^2},~\omega_{\pm} = [\text{cosh}(\beta \mathcal{N}) \pm \mathfrak{W}],\label{eq2.26} \\  
    \text{with,}~\mathfrak{W} = \sqrt{\sinh^2{[\frac{\beta}{2} (\mathbb{B}\sin^2{\gamma}-\mathbb{C}\cos^2{\gamma})]} + 1},\nonumber\\
    \text{and}~~\mathcal{N} = \frac{1}{2}(\mathbb{B}\sin^2{\gamma}-\mathbb{C}\cos^2{\gamma}).  \label{eq2.27}
\end{gather}
\textbf{4. Player's payoff average:} The analytical expression for $\langle \Lambda \rangle^{NEM}$ in QuPD can be derived using $\Upsilon^{NEM}$ given in Eq.~(\ref{eq2.18}) as,
\begin{gather}
    \Upsilon^{NEM} = e^{\beta\mathrm{(m-p)}} + e^{-\beta \mathrm{(n+q)}} + 2e^{\frac{\beta}{2} \mathrm{(n+p-m-q)}},\nonumber\\
    \text{or,}~\Upsilon^{NEM} = e^{\beta(\mathbb{B}\sin^2{\gamma} - \mathbb{C}\cos^2{\gamma})} + e^{\beta ( \mathbb{C}\cos^2{\gamma} - \mathbb{B}\sin^2{\gamma} )} + 1.\nonumber\\
    \text{Thus,}~~\langle {\Lambda} \rangle^{NEM} = -\dfrac{1}{2}\langle \mathbb{E}\rangle^{NEM} = \dfrac{1}{2}\bigg[\dfrac{1}{\Upsilon^{NEM}}\dfrac{\partial \Upsilon^{NEM}}{\partial \beta} \bigg], \nonumber\\
    \text{or,}~\langle {\Lambda} \rangle^{NEM}= \dfrac{\mathcal{N} (e^{2\beta\mathcal{N}} - e^{-2\beta\mathcal{N}})}{(1 + e^{2\beta\mathcal{N}} + e^{-2\beta\mathcal{N}})},
    \label{eq2.28}
\end{gather}
where, $\langle \mathbb{E}\rangle$ denotes the average internal energy (\textbf{NOTE}: we take the $-ve$ of $\langle \mathbb{E}\rangle$ to maximize $\langle {\Lambda} \rangle$) and $\mathcal{N} = \frac{1}{2}(\mathbb{B}\sin^2{\gamma}-\mathbb{C}\cos^2{\gamma})$, respectively.  
\\
\textbf{5. Payoff capacity:} The analytical expression for $\wp_{C}^{NEM}$ in QuPD can also be derived from $\Upsilon^{NEM}$ given in Eqs.~(\ref{eq2.18}, \ref{eq2.20}, \ref{eq2.28}) as,
\begin{gather}
    \Upsilon^{NEM} = e^{\beta(\mathbb{B}\sin^2{\gamma} - \mathbb{C}\cos^2{\gamma})} + e^{\beta ( \mathbb{C}\cos^2{\gamma} - \mathbb{B}\sin^2{\gamma} )} + 1.\nonumber
\end{gather}
Since the payoff capacity $\wp_C$ is analogous to the normalized thermodynamic specific heat capacity, at constant volume, $\complement_{\mathbb{V}}$ (see, Ref.~\cite{ref12, ref14, ref21, ref22, ref23}), we have the analytical NEM expression for $\wp_C$, normalized by $\beta^2$, as,
\begin{gather}
    \wp_{C}^{NEM} = \dfrac{1}{2\beta^2} \dfrac{\partial}{\partial\beta} \bigg[\dfrac{1}{\Upsilon^{NEM}}\dfrac{\partial \Upsilon^{NEM}}{\partial \beta} \bigg],\nonumber\\
    \text{or,}~~\wp_{C}^{NEM} = 2\mathcal{N}^2\dfrac{e^{2\beta\mathcal{N}} (1+ 4e^{2\beta\mathcal{N}} + e^{4\beta\mathcal{N}})}{(1 + e^{2\beta\mathcal{N}} + e^{4\beta\mathcal{N}})^2},
    \label{eq2.29}
\end{gather}
where, $\mathcal{N} = \frac{1}{2}(\mathbb{B}\sin^2{\gamma}-\mathbb{C}\cos^2{\gamma})$. We will study the variation of these \textit{five} indicators with respect to a changing $\gamma\in [0, \pi]$ while keeping $\mathbb{B} = 5.0$ and $\mathbb{C} = 2.0$ (i.e., constant values). In the next section, we will discuss the algorithm for Agent-based modelling.

\subsubsection{\label{ABM}Agent-based Modelling (ABM)}
ABM \cite{ref10, ref12} is a numerical modelling technique often used to study classical SD in the \textit{TL}. However, to the best of our knowledge, ABM has not been previously implemented to study the emergence of cooperative behaviour in quantum social dilemmas like QuPD, etc. Hence, the main attraction of our work is that we numerically analyze the emergence of cooperation among an infinite number of quantum players, in the presence of entanglement $\gamma$. We consider a \textit{1000} quantum players, and these quantum players reside on the $1D$-IC where each site consists of two entangled quantum players, playing the $two$-strategy QuPD, and they interact with their nearest neighbours only, in the presence of a periodic boundary condition. The energy matrix $\Delta$ is just the negative of the QuPD payoff matrix $\Lambda$ given in Eq.~(\ref{eq2.11}), and this gives the IC's individual site energy. At this point, we modify the quantum player's strategy by iterating through a conditional loop \textit{1,000,000} times, which amounts to an average of \textit{1000} strategy modifications per quantum player. While Refs.~\cite{ref10, ref12} provide a clear explanation of the algorithm's basic structure (which is based on the \textit{Metropolis algorithm}\cite{ref21, ref22}), we still need to figure out the game magnetization, entanglement susceptibility, correlation, player's payoff average, and payoff capacity for our particular scenario. Ref.~\cite{jan} studies QuPD game with around 2000 players on scale free or random networks, however it is not on an Ising chain and neither is it based on the basic structure of the metropolis algorithm. 

So, this is a quick synopsis of the ABM algorithm:
\begin{enumerate}
    \item At each site on the $1D$-IC, assign a random strategy to each quantum player: \textit{0} (strategy $\$_1$: say \textit{defection}) or \textit{1} (strategy $\$_2$: say \textit{quantum}).
    \item Choose a principal quantum player at random to determine both its unique strategy as well as the strategy of its closest neighbour. The energy $\Delta$ of the principal quantum player is ascertained based on the strategies that have been determined. The principal quantum player's energy is computed for each of the two scenarios: either it chose the opposite strategy while preserving the closest neighbour's strategy, or it chose the same strategy as its closest neighbour. 
    \item For each of the two possible outcomes, the energy difference ($d\Delta$) is determined for the principal quantum player. The current strategy of the principal quantum player is flipped based on whether the Fermi function $(1 + e^{\beta \cdot d\Delta})^{-1} > 0.5$; if not, it is not flipped.~\cite{ref21}. 
    \item Now, based on the indicators, five distinct conditions emerge:
    \begin{itemize}
        \item \textit{Game magnetization}: After each run of the conditional spin-flipping loop, we calculate the difference between the number of quantum players playing \textit{quantum} (or, $\mathbb{Q}$) and the number of quantum players playing \textit{defect} (or, $\mathfrak{D}$). This gives the total magnetization $\Tilde{\mu}= \sum_{i} \sigma_i$, for $\sigma_i$ being the strategy ($0$ or $1$) of the quantum player at the $i^{th}$-site, for each cycle of the conditional loop. Finally, we take the average of the total magnetization for all the loops to determine the overall game magnetization, i.e., $\mu^{ABM} = \langle \Tilde{\mu}  \rangle$.
        \item \textit{Entanglement susceptibility}: Following Eqs. (\ref{eq2.20},\ref{eq2.22}), the entanglement susceptibility can be determined from the variance of $\mu^{ABM}$ as, $\chi_\gamma^{ABM}= \frac{(\mathbb{B} + \mathbb{C})}{2} \sin{(2\gamma)} [\langle \Tilde{\mu}^2 \rangle - \langle \Tilde{\mu} \rangle^2]$, where, $\Tilde{\mu} = \sum_{i} \sigma_i$ and $\mu^{ABM} = \langle \Tilde{\mu}  \rangle$, respectively. Hence, following the previous steps for determining $\mu^{ABM}$, we compute the $\mu^{ABM}$ variance and multiply the result by the factor: $\frac{1}{2}(\mathbb{B}+\mathbb{C})\sin{(2\gamma)}$, to get the entanglement susceptibility.
        \begin{figure*}[!ht]
    \centering
    \begin{subfigure}[b]{0.95\columnwidth}
        \centering
        \includegraphics[width = \textwidth]{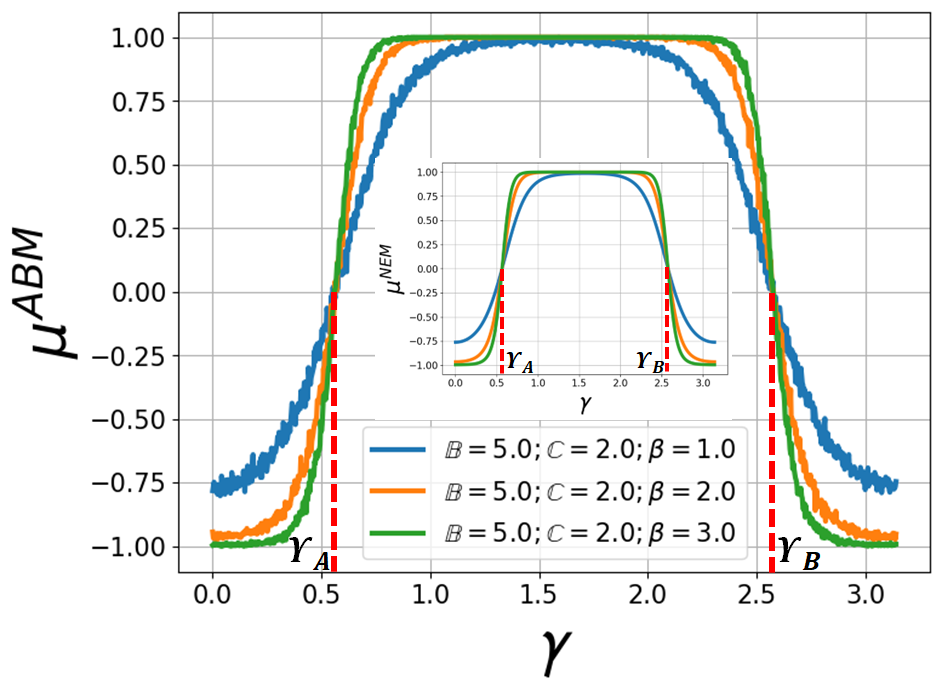}
        \caption{$\mu^{ABM/NEM}$ vs $\gamma$}
        \label{fig1a}
    \end{subfigure}
    \begin{subfigure}[b]{0.95\columnwidth}
        \centering
        \includegraphics[width = \textwidth]{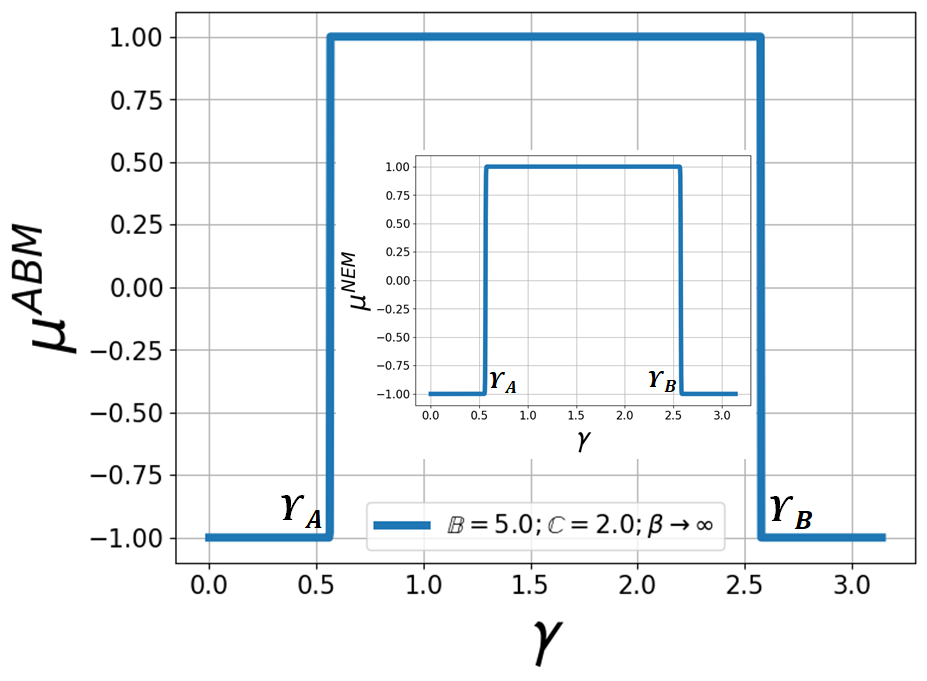}
        \caption{$\mu^{ABM/NEM}$ vs $\gamma$ in $\beta\rightarrow \infty$ (or, $T\rightarrow 0$) limit}
        \label{fig1b}
    \end{subfigure}
    \caption{\centering{ABM and NEM (in \textit{insets}): \textbf{Game magnetization} $\mu$ vs \textbf{entanglement} $\gamma$ (a) for $ \beta=1, 2, 3$ and (b)in the limit $\beta \rightarrow \infty$. Other parameters common to both are \textbf{reward} $\mathbb{R}= (\mathbb{B-C}) = 3.0$, \textbf{sucker's payoff} $\mathbb{S}= \mathbb{-C}=-2.0$, \textbf{temptation} $\mathbb{T} = \mathbb{B} =5.0$, \textbf{punishment} $\mathbb{P}=0.0$, $\gamma_A = 0.5639$ and $\gamma_B=2.5777$ in QuPD.}}
    \label{fig:1}
\end{figure*} 
        \item \textit{Correlation}: We take into account two entangled principal quantum player pairs, thus we slightly alter the first four steps to incorporate the needs of determining the correlation ($\mathfrak{c}_j^{ABM}$) between the two pairs of entangled quantum players. Following the individual spin-flipping operations for the two randomly selected principal quantum player pairs, the total correlation is increased by a correlation value of $+1$ if both quantum players' strategies are the same and a correlation value of $-1$ if they are different. 
        \item \textit{Player's payoff average}: Similar to the case of previous indicators, after all the spin-flipping operations, we determine the total energy of each individual Ising site, and the $-ve$ of this energy gives us the player's payoff average (or, $\langle\Lambda\rangle^{ABM}$).
        \item \textit{Payoff capacity}: Similar to $\chi_\gamma^{ABM}$, the $\wp_C^{ABM}$ can be determined from the variance of average internal energy $\langle \mathbb{E} \rangle$, or, $-\langle\Lambda\rangle^{ABM}$ (see, Eqs.~(\ref{eq2.28}, \ref{eq2.30}) and Ref.~\cite{ref12}), and this gives us, \begin{equation}
        \wp_{C}^{ABM} =\dfrac{1}{2}[\langle\mathbb{E}^2\rangle - \langle\mathbb{E}\rangle^2].
        \label{eq2.30} \end{equation}
        Hence, following the previous steps for determining the player's payoff average, we calculate the variance of the internal energy (or, player's payoff average) and multiply it with the factor $\frac{1}{2}$ to get the payoff capacity.
    \end{itemize} 
    \item Proceed to step 2 and carry out this procedure a \textit{1000} times.
\end{enumerate}
After reviewing the Python codes included in Appendices~(\ref{appendix1}, \ref{appendix2}), one can have a better understanding of this algorithm. Given that our primary goal is to maximise the feasible payoff — which we can only do when our system reaches the energy equilibrium or the least energy configuration — we see that the likelihood of strategy switching drops as the energy difference $d\Delta$ increases.

\section{\label{rsa}Results and Analysis}
In our version of QuPD, there are effectively 2 game payoffs: Cooperation bonus ($\mathbb{B}$) and the Cost ($\mathbb{C}$), along with the measure of Entanglement ($\gamma$), with $\mathbb{B}\geq \mathbb{C}\geq 0$. We also restrict the value of $\gamma \in [0, \pi]$, respectively. When $\gamma = \frac{\pi}{2}$, we observe maximal entanglement among the quantum players, and this signifies a \textit{Bell state}. Here, we will analyze the variation of all five indicators: Game magnetization, Entanglement susceptibility, Correlation, Player's payoff average and Payoff capacity, with respect to the Entanglement $\gamma$ while $\mathbb{B} = 5.0$ and $\mathbb{C} = 2.0$. One thing to note is that entanglement susceptibility, compared to game magnetization, is a far more accurate tool for gauging minute alterations in the number of quantum players playing \textit{defect} or \textit{quantum}, owing to a change in entanglement $\gamma$.   

\subsection{Game magnetization $(\mu)$}
\subsubsection{\underline{NEM}}
Using the given values of $\mathbb{B} = 5.0$, $\mathbb{C} = 2.0$ and the expression of $\mu^{NEM}$ in Eq.~(\ref{eq2.21}), we plot the variation of game magnetization with regard to changing values of $\gamma$, and they are given in the \textit{insets} of Figs.~\ref{fig1a}, \ref{fig1b}. We observe a change in the sign of the game magnetization from negative to positive and vice-versa at two particular values of $\gamma$. These indicate \textit{first}-order phase transitions in the strategies adopted by the quantum players, changing from \textit{defect} to \textit{quantum} and vice-versa, respectively. To determine the two critical values of $\gamma$ (say, $\gamma_A$ and $\gamma_B$) where these phase transitions occur, we equate $\mu^{NEM} = 0$ (see, Eq.~(\ref{eq2.21})) and this leads to $\frac{(\mathbb{B}\sin^2{\gamma}-\mathbb{C}\cos^2{\gamma})}{2} = 0$, finally giving us the relation between the critical $\gamma$ values and the game payoffs ($\mathbb{B}, \mathbb{C}$) as: $\gamma_{A,B} = \tan^{-1}\sqrt{\mathbb{C}/\mathbb{B}}$. For $\mathbb{B} = 5.0$ and $\mathbb{C} = 2.0$, we have $\gamma = \tan^{-1}\sqrt{{2}/{5}} = 0.5639~\text{or}~2.5777$, i.e., $\gamma_A = 0.5639$ and $\gamma_B = 2.5777$, respectively (see, Figs.~\ref{fig1a}, \ref{fig1b}) and they are independent of the \textit{noise} $\beta$. Thus, the classical \textit{defect} phase appears in the regime $\gamma \in [0, \gamma_A) \cup (\gamma_B,\pi]$ while the \textit{quantum} phase appears in the regime $\gamma\in [\gamma_A, \gamma_B]$. For any general case, if $\mathbb{B}\gg \mathbb{C}$, then $\gamma_{A,B} = \tan^{-1}\sqrt{\mathbb{C}/\mathbb{B}} \rightarrow 0~\text{or}~\pi$, and this indicates that the classical \textit{defect} ($\mathfrak{D}$) phase disappears and the \textit{quantum} phase is the only phase. By definition (see, CPD in Sec.~\ref{CPD}), $\mathbb{B}$ is always greater than $\mathbb{C}$, hence, there always exists a quantum phase, otherwise, if $\mathbb{B}=\mathbb{C}$ or $\mathbb{B}<\mathbb{C}$, then the prisoner's dilemma disappears. However, when $\mathbb{B} \sim \mathbb{C}$, $\gamma_{A,B} = \tan^{-1}\sqrt{\mathbb{C}/\mathbb{B}} \approx \tan^{-1}(\pm 1) \rightarrow \frac{\pi}{4}$ or $\frac{3\pi}{4}$, indicating that the classical \textit{defect} phase appears in the region $\gamma \in [0, \frac{\pi}{4}) \cup (\frac{3\pi}{4}, \pi]$ while the \textit{quantum} phase appears in the region $\gamma \in [\frac{\pi}{4}, \frac{3\pi}{4}]$.

In the Z-N limit, i.e., $T\rightarrow 0$ (or, $\beta \rightarrow \infty$), if $\gamma$ lies between $\gamma_A$ and $\gamma_B$, then $\mu^{NEM} \rightarrow 1,~\forall~\gamma_A < \gamma < \gamma_B$, indicating that all quantum players play the \textit{quantum} strategy. However, when $\gamma$ is either lower than $\gamma_A$ or greater than $\gamma_B$, i.e., $\gamma \in [0, \gamma_A) \cup (\gamma_B, \pi]$, in the same Z-N limit, $\mu^{NEM} \rightarrow -1,~\forall~\gamma \in [0, \gamma_A) \cup (\gamma_B, \pi]$, indicating that all quantum players play the \textit{defect} strategy. In the I-N limit, i.e., $T\rightarrow \infty$ (or, $\beta \rightarrow 0$), $\mu^{NEM} \rightarrow 0$, implying that the quantum players opt for their strategies randomly, resulting in an equiprobable number selecting \textit{defect} and \textit{quantum}, respectively.

\subsubsection{\label{e1}\underline{ABM}}
For the given values of $\mathcal{T} = 0$ and $\mathcal{F} = \frac{(\mathbb{B}\sin^2{\gamma}-\mathbb{C}\cos^2{\gamma})}{2}$ in Eq.~(\ref{eq2.20}), we have the Energy matrix $\Delta = -{\Lambda}$ (see, Eq.~(\ref{eq2.11})). Thus,
\begin{equation}
    \Delta = \left[\begin{array}{c c}  
         -(\mathbb{B-C}) & -( \mathbb{B}\sin^2{\gamma}-\mathbb{C}\cos^2{\gamma})\\
         -(\mathbb{B}\cos^2{\gamma} - \mathbb{C}\sin^2{\gamma}) & 0
    \end{array}\right].
        \label{eq3.1}
\end{equation}
Following the algorithm given in Sec.~\ref{ABM}, for $\mathbb{B} = 5.0$ and $\mathbb{C} = 2.0$, we determine the ABM game magnetization ($\mu^{ABM}$), and its variation with the entanglement $\gamma$ is shown in Figs.~\ref{fig1a}, \ref{fig1b}. We observe exactly the same results obtained via both NEM and ABM in the finite and limiting values of $\beta$. For $\beta>0$, when $\gamma \in [0, \gamma_A)\cup (\gamma_B, \pi]$, the majority of quantum players play \textit{defect} over \textit{quantum}. However, for the same $\beta>0$, when $\gamma \in [\gamma_A, \gamma_B]$, the \textit{quantum} phase appears, and a majority of quantum players play \textit{quantum}. 

In the Z-N limit, i.e., $T\rightarrow 0$ (or, $\beta \rightarrow \infty$), $\mu^{ABM} \rightarrow 1,~\forall~\gamma \in [\gamma_A , \gamma_B]$, indicating that all quantum players play the \textit{quantum} strategy. However, when $\gamma \in [0, \gamma_A) \cup (\gamma_B, \pi]$, in the same Z-N limit, $\mu^{ABM} \rightarrow -1,~\forall~\gamma \in [0, \gamma_A) \cup (\gamma_B, \pi]$, indicating that all quantum players play the \textit{defect} strategy. In the I-N limit, i.e., $T\rightarrow \infty$ (or, $\beta \rightarrow 0$), $\mu^{ABM} \rightarrow 0$, implying that the quantum players oft for their strategies randomly, resulting in an equiprobable number selecting \textit{defect} and \textit{quantum}, respectively.

\subsubsection{\underline{Analysis of game magnetization}}
The analysis of game magnetization: $\mu^{NEM}$ as well as $\mu^{ABM}$, will be done in this subsection. We observe in Fig.~\ref{fig:1} that when $\gamma\rightarrow\frac{\pi}{2}$, i.e., \textit{maximal entanglement}, $\mu^{ABM} = \mu^{NEM} \rightarrow 1$, indicating that all quantum players play the \textit{quantum} strategy. However, when $\gamma\rightarrow 0~\text{or}~\pi$, i.e., \textit{zero entanglement}, a large fraction of quantum players play the classical \textit{defect} strategy. For both finite and limiting $\beta$ values, we observe \textit{first}-order phase transitions at the two critical values of $\gamma$ (i.e., at $\gamma_A$ and $\gamma_B$) and the values of $\gamma_{A, B}$ depends on the payoffs $\mathbb{B}$ and $\mathbb{C}$ via the relation: $\gamma_{A, B} = \tan^{-1}\sqrt{\mathbb{C}/\mathbb{B}}$ (see, Sec.~\ref{e1}). 
\begin{figure}[!ht]
    \centering
    \includegraphics[width = \columnwidth]{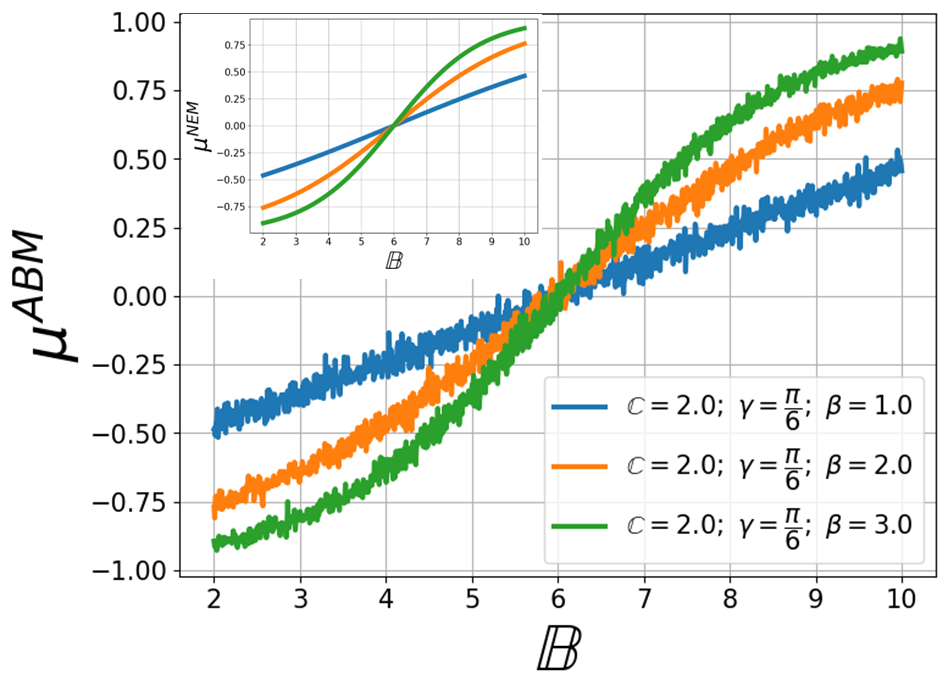}
    \caption{\centering{\centering{ABM and NEM (in \textit{insets}): $\mu^{ABM/NEM}$ vs changing \textbf{cooperation bonus} $\mathbb{B}$ for fixed \textbf{cost} $\mathbb{C}=2.0$ and \textbf{entanglement} $\gamma = \frac{\pi}{6}$ in QuPD.}}}
    \label{fig:0a}
\end{figure}
For fixed values of $\gamma$ (say, $\gamma=\frac{\pi}{6}$) and $\mathbb{C}$ (say, $\mathbb{C} = 2.0$), if we vary $\mathbb{B}$ from $\mathbb{B} \sim \mathbb{C}$ to $\mathbb{B}\gg \mathbb{C}$, while satisfying the criterion: $\mathbb{B}>\mathbb{C}$, we observe a phase transition from \textit{defect} to \textit{quantum} as the cooperation bonus, i.e., $\mathbb{B}$ increases, and the phase transition (\textit{PT}) occurs at: $\mathbb{B}_{PT} = \mathbb{C}\cot^2{\gamma}$ (see, Fig.~\ref{fig:0a}). Similarly, for fixed $\gamma$ and $\mathbb{B}$, if we vary $\mathbb{C}$ from $\mathbb{C} \ll \mathbb{B}$ to $\mathbb{C}\sim \mathbb{B}$, while satisfying the criterion: $\mathbb{C}<\mathbb{B}$, we observe a phase transition from \textit{quantum} to \textit{defect} as the cost associated with the game, i.e., $\mathbb{C}$ increases, and the phase transition occurs at: $\mathbb{C}_{PT} = \mathbb{B}\tan^2{\gamma}$. 
\begin{figure*}[!ht]
    \centering
    \begin{subfigure}[b]{0.95\columnwidth}
        \centering
        \includegraphics[width = \textwidth]{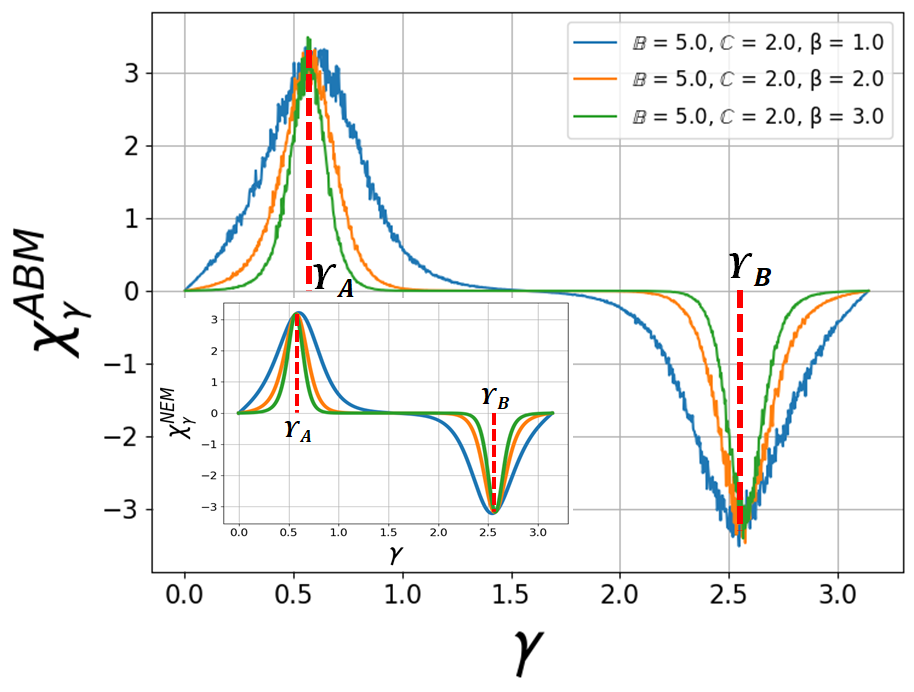}
        \caption{$\chi_\gamma^{ABM/NEM}$ vs $\gamma$}
        \label{fig2a}
    \end{subfigure}
    \begin{subfigure}[b]{0.95\columnwidth}
        \centering
        \includegraphics[width = \textwidth]{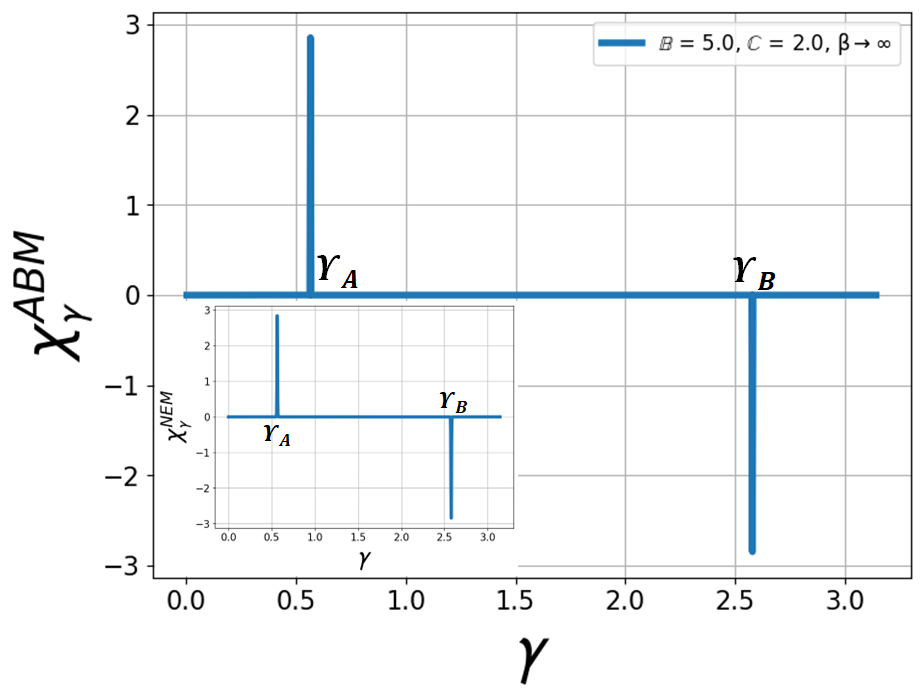}
        \caption{$\chi_\gamma^{ABM/NEM}$ vs $\gamma$ in $\beta\rightarrow \infty$ (or, $T\rightarrow 0$) limit}
        \label{fig2b}
    \end{subfigure}
    \caption{\centering{ABM and NEM (in \textit{insets}): \textbf{Entanglement susceptibility} $\chi_\gamma$ vs \textbf{entanglement} $\gamma$ (a) for $\beta=1, 2, 3$  and (b) for $\beta\rightarrow \infty$. Other parameters common to both are: \textbf{reward} $\mathbb{R}= (\mathbb{B-C}) =3.0$, \textbf{sucker's payoff} $\mathbb{S}= \mathbb{-C}= -2.0$, \textbf{temptation} $\mathbb{T}= \mathbb{B}=5.0$, \textbf{punishment} $\mathbb{P}=0.0$, $\gamma_A = 0.5639$ and $\gamma_B=2.5777$ in QuPD.}}
    \label{fig:2}
\end{figure*} 
Hence, the game payoffs, i.e., $(\mathbb{B},\mathbb{C})$ can also induce phase transition as seen from Fig.~\ref{fig:0a}. Thus, we can have the phase transition(s) occurring in QuPD via both game payoffs and entanglement.

\subsection{Entanglement susceptibility $(\chi_\gamma)$}
\subsubsection{\underline{NEM}}
Using the given values of $\mathbb{B} = 5.0$, $\mathbb{C} = 2.0$ and the expression of $\chi_{\gamma}^{NEM}$ in Eq.~(\ref{eq2.22}), we plot the variation of entanglement susceptibility with regard to changing values of $\gamma$ and they are shown in the \textit{insets} of Figs.~\ref{fig2a}, \ref{fig2b}. Here, in the $\beta\rightarrow\infty$ (i.e., Z-N) limit, we observe \textit{two} sharp discontinuous peaks at two values of $\gamma$. To determine these critical $\gamma$ values where $\chi^{NEM}_{\gamma}\rightarrow\infty$, in the $\beta\rightarrow\infty$ limit, we equate $\frac{1}{\chi_{\gamma}^{NEM}}\rightarrow 0$ (see, Eq.~(\ref{eq2.22})) and we get the condition: $\mathcal{N} = \frac{1}{2}(\mathbb{B}\sin^2{\gamma}-\mathbb{C}\cos^2{\gamma}) = 0$, and this gives us the same expression for the critical $\gamma$'s as in the case of game magnetization, i.e., $\gamma_{A, B} = \tan^{-1}\sqrt{\mathbb{C}/\mathbb{B}}$. For $\mathbb{B} = 5.0$ and $\mathbb{C} = 2.0$, we have $\gamma = \tan^{-1}\sqrt{{2}/{5}} = 0.5639~\text{or}~2.5777$. Therefore, we have $\gamma_A = 0.5639$ and $\gamma_B = 2.5777$, for the given values of $\mathbb{B} = 5.0$ and $\mathbb{C} = 2.0$ (see, Figs.~\ref{fig2a}, \ref{fig2b}) and they are independent of the \textit{noise} $\beta$. 

When $\beta$ is finite, as $\gamma \in [0, \gamma_A]$ gradually increases, we note that the strategy switching rate from \textit{defect} to \textit{quantum} also increases, and this rate peaks at $\gamma = \gamma_A$. Then, as we increase the value of $\gamma$ from $\gamma_A$ to $\gamma = \frac{\pi}{2}$, the switching rate decreases, even though \textit{quantum} strategy still remains the dominant strategy. At maximal entanglement, i.e., $\gamma = \frac{\pi}{2}$, the entanglement susceptibility vanishes since all the quantum players play \textit{quantum}, and they do not shift to \textit{defect} on minute changes in $\gamma$. However, as we further increase the $\gamma$ value from $\frac{\pi}{2}$ to $\gamma_B$, the strategy switching rate from \textit{quantum} to \textit{defect} increases, peaking at $\gamma_B$, and we observe that \textit{defect} gradually becomes the dominant strategy of most quantum players. For $\gamma=0~\text{or}~\pi$, a large fraction of quantum players play \textit{defect}. In the Z-N limit, i.e., $T\rightarrow 0$ (or, $\beta \rightarrow \infty$), $\chi_{\gamma}^{NEM} \rightarrow 0,~\forall~\gamma\neq \{\gamma_A, \gamma_B\}$ and here we observe the two sharp discontinuous peaks at $\gamma = \gamma_A$ and $\gamma=\gamma_B$ as discussed before. They indicate two \textit{first}-order phase transitions at $\gamma_A$ and $\gamma_B$ since the game magnetization plot shows a discontinuity at $\gamma_A$ and $\gamma_B$ in the Z-N limit (see, Fig.~\ref{fig1b}). The \textit{first}-order phase transition is also a characteristic of \textit{Type-I} superconductors (below a certain critical temperature and in the absence of an external field), and this shows that entanglement plays an important role in exhibiting phase transitions in quantum games like QuPD. In Z-N limit, for $\gamma\neq \{\gamma_A, \gamma_B\}$, the $\mu^{NEM} \rightarrow \pm 1$ and hence they do not change with a changing $\gamma$, resulting in $\chi_{\gamma}^{NEM}\rightarrow 0,~\forall~\gamma\neq \{\gamma_A, \gamma_B\}$.

In the I-N limit, i.e., $T\rightarrow \infty$ (or, $\beta \rightarrow 0$), $\chi_{\gamma}^{NEM} \rightarrow \frac{(\mathbb{B} + \mathbb{C})}{2} \sin{(2\gamma)},~\forall~\gamma$, implying that the quantum players opt for their strategies randomly. The interesting part to discuss here is the unique expression of $\lim_{\beta\rightarrow 0}\chi_{\gamma}^{NEM} \rightarrow \frac{(\mathbb{B} + \mathbb{C})}{2} \sin{(2\gamma)},~\forall~\gamma$ in the I-N limit. In the I-N limit, Taylor expanding the expression of $\mu^{NEM}$, in Eq.~(\ref{eq2.21}), around $\beta$, makes this very easily verifiable, where the $1^{st}$-order correction (say, $\mu^{(1)}$) of $\mu^{NEM}$, is given as, $\mu^{(1)} = -\frac{\beta(\mathbb{B} + \mathbb{C})}{4}\cos{(2\gamma)}$, leading to the $0^{th}$-order $\chi_\gamma^{NEM}$ correction as,
\begin{equation}
    \lim_{\beta\rightarrow 0}\chi_{\gamma}^{(0)} = \lim_{\beta\rightarrow 0}\frac{1}{\beta} \frac{\partial}{\partial \gamma}\mu^{(1)} =  \frac{(\mathbb{B} + \mathbb{C})}{2} \sin{(2\gamma)}.
    \label{eq3.3}
\end{equation}
The \textit{higher} order expansion terms of $\chi_\gamma^{NEM}$ about $\beta$, in the I-N limit, vanishes. This explains why we get a non-zero expression for $\chi_{\gamma}$ in the I-N limit. 
\subsubsection{\underline{ABM}}
For the given values of $\mathcal{T} = 0$ and $\mathcal{F} = \frac{(\mathbb{B}\sin^2{\gamma}-\mathbb{C}\cos^2{\gamma})}{2}$ in Eq.~(\ref{eq2.20}), we have $\Delta = -{\Lambda}$ (see, Eq.~(\ref{eq2.11})). Thus,
\begin{equation}
    \Delta = \left[\begin{array}{c c}  
         -(\mathbb{B-C}) & -( \mathbb{B}\sin^2{\gamma}-\mathbb{C}\cos^2{\gamma})\\
         -(\mathbb{B}\cos^2{\gamma} - \mathbb{C}\sin^2{\gamma}) & 0
    \end{array}\right].
        \label{eq3.2}
\end{equation}
Following the algorithm given in Sec.~\ref{ABM}, for $\mathbb{B} = 5.0$ and $\mathbb{C} = 2.0$, we determine the ABM entanglement susceptibility ($\chi_{\gamma}^{ABM}$), and its variation with the entanglement $\gamma$ is shown in Figs.~\ref{fig2a}, \ref{fig2b}. Here also, we observe exactly the same results as obtained for entanglement susceptibility, via NEM, in the finite and limiting values of $\beta$. For finite $\beta$ and increasing values of $\gamma \in [0, \gamma_A]$, we note that the strategy switching rate from \textit{defect} to \textit{quantum} also increases, reaching the peak value at $\gamma = \gamma_A$. On further increase of $\gamma$ value, from $\gamma_A$ to $\gamma = \frac{\pi}{2}$, the switching rate decreases, even though \textit{quantum} still remains the dominant strategy. In the case of maximal entanglement, i.e., $\gamma = \frac{\pi}{2}$, $\chi_{\gamma}^{ABM}$ vanishes since all the quantum players play the \textit{quantum} strategy, and they do not change their strategies on minute changes in $\gamma$. On further increase of $\gamma$ value, from $\frac{\pi}{2}$ to $\gamma_B$, the strategy switching rate from \textit{quantum} to \textit{defect} increases, peaking at $\gamma_B$, and we observe that \textit{defect} gradually becomes the dominant strategy of most quantum players. For $\gamma=0~\text{or}~\pi$, a large fraction of quantum players play \textit{defect}.

In the Z-N limit, i.e., $T\rightarrow 0$ (or, $\beta \rightarrow \infty$), $\chi_{\gamma}^{ABM} \rightarrow 0,~\forall~\gamma\neq \{\gamma_A, \gamma_B\}$. Here also, we observe two sharp discontinuous peaks, at $\gamma = \gamma_A$ and $\gamma=\gamma_B$, indicating the two \textit{first}-order phase transitions at $\gamma_A$ and $\gamma_B$. As seen in Fig.~\ref{fig1b}, the game magnetization plot shows a discontinuity at $\gamma_A$ and $\gamma_B$ in the Z-N limit. In Z-N limit, for $\gamma\neq \{\gamma_A, \gamma_B\}$, the $\mu^{ABM} \rightarrow \pm 1$ and hence they do not change with a changing $\gamma$, resulting in $\chi_{\gamma}^{ABM}\rightarrow 0,~\forall~\gamma\neq \{\gamma_A, \gamma_B\}$. In the I-N limit, i.e., $T\rightarrow \infty$ (or, $\beta \rightarrow 0$), $\chi_{\gamma}^{ABM} \rightarrow \frac{(\mathbb{B} + \mathbb{C})}{2} \sin{(2\gamma)},~\forall~\gamma$, implying that the quantum players opt for their strategies randomly. In the I-N limit, even though $\langle\mu^{ABM}\rangle \rightarrow 0$ due to strategy selection randomization,  $\langle(\mu^{ABM})^2\rangle \rightarrow 1$ and this gives us the value of $\chi_{\gamma}^{ABM} = \frac{(\mathbb{B} + \mathbb{C})}{2} \sin{(2\gamma)}$ (see, Eq.~(\ref{eq2.22})) in the $\beta\rightarrow 0$ (or, I-N) limit.

\subsubsection{\underline{Analysis of entanglement susceptibility}}
The analysis of entanglement susceptibility: $\chi_\gamma^{NEM}$ as well as $\chi_\gamma^{ABM}$, will be done in this subsection. From Fig.~\ref{fig:2}, we observe that for all $\gamma \rightarrow \{0,~\frac{\pi}{2},~\pi\}$, i.e., for both \textit{minimal} (i.e., $\gamma\rightarrow 0~\text{or}~\pi$) and \textit{maximal entanglement} (i.e., $\gamma\rightarrow\frac{\pi}{2}$), $\chi_\gamma^{ABM} = \chi_\gamma^{NEM} \rightarrow 0$, indicating that there is no phase transition among the quantum players. This can also be verified from the game magnetization result (see, Fig.~\ref{fig:1}) where we observe that for \textit{finite} as well as \textit{limiting} values of $\beta$, when $\gamma \rightarrow \{0,~\frac{\pi}{2},~\pi\}$, a large fraction of quantum players play either \textit{defect} (for $\gamma\rightarrow0$ or $\pi$) or \textit{quantum} (for $\gamma\rightarrow\frac{\pi}{2}$), and this leads to a vanishing $\chi_\gamma$. Interestingly, in the Z-N (or, $\beta\rightarrow\infty$) limit, we observe two \textit{first}-order phase transitions, as shown in Fig.~\ref{fig2b}, at the two critical values of $\gamma$ (i.e., at $\gamma_A$ and $\gamma_B$). The values of $\gamma_{A, B}$, where we observe the phase transitions, depend on the payoffs $\mathbb{B}$ and $\mathbb{C}$ via the relation: $\gamma_{A, B} = \tan^{-1}\sqrt{\mathbb{C}/\mathbb{B}}$ (see, Sec.~\ref{e1}). The two \textit{first}-order phase transition points, i.e., $\gamma_A$ and $\gamma_B$, depend on both game payoffs (i.e., $\mathbb{B},~\mathbb{C}$), indicating that the game payoffs can also induce phase transition as seen in the case of $\mu$. 
\begin{figure*}[!ht]
    \centering
    \begin{subfigure}[b]{0.95\columnwidth}
        \centering
        \includegraphics[width = \textwidth]{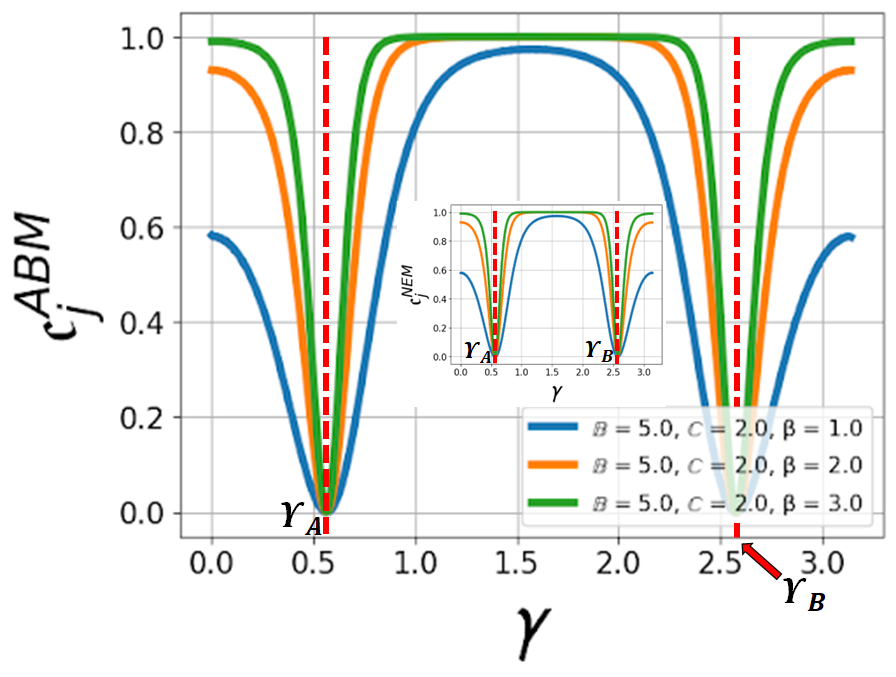}
        \caption{$\mathfrak{c}_j^{ABM/NEM}$ vs $\gamma$}
        \label{fig3a}
    \end{subfigure}
    \begin{subfigure}[b]{0.95\columnwidth}
        \centering
        \includegraphics[width = \textwidth]{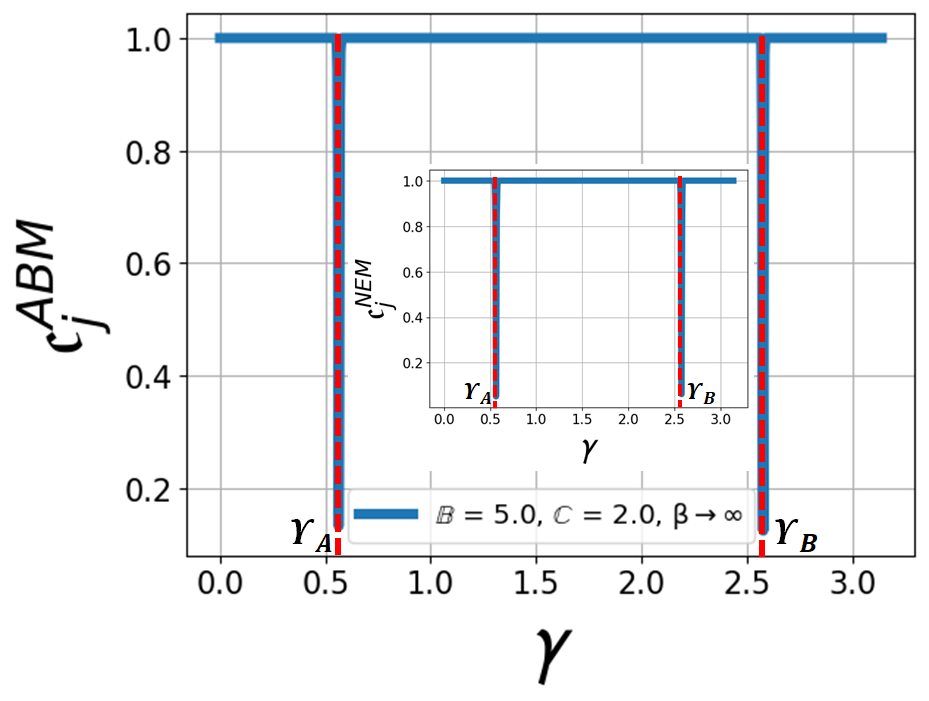}
        \caption{$\mathfrak{c}_j^{ABM/NEM}$ vs $\gamma$ in $\beta\rightarrow \infty$ (or, $T\rightarrow 0$) limit}
        \label{fig3b}
    \end{subfigure}
    \caption{\centering{ABM and NEM (in \textit{insets}): \textbf{Correlation} $\mathfrak{c}_j$ vs $\gamma$ for (a) $\beta=1, 2, 3 $ and (b) $\beta\rightarrow \infty$. Other parameters common to both are \textbf{distance} $j=11$, \textbf{reward} $\mathbb{R}= (\mathbb{B-C}) =3.0$, \textbf{sucker's payoff} $\mathbb{S}= \mathbb{-C}= -2.0$, \textbf{temptation} $\mathbb{T}= \mathbb{B}=5.0$, \textbf{punishment} $\mathbb{P}=0.0$, $\gamma_A = 0.5639$ and $\gamma_B=2.5777$ in QuPD.}}
    \label{fig:3}
\end{figure*}
\subsection{Correlation $(\mathfrak{c}_j)$}
\subsubsection{\underline{NEM}}
Using the given values of $\mathbb{B} = 5.0$, $\mathbb{C} = 2.0$ and the expression of $\mathfrak{c}_j^{NEM}$ in Eqs.~(\ref{eq2.25}, \ref{eq2.26}, \ref{eq2.27}), we plot the variation of correlation with regard to changing values of $\gamma$ and they are shown in the \textit{insets} of Figs.~\ref{fig3a}, \ref{fig3b}. In both Figs.~\ref{fig3a} and \ref{fig3b}, we observe a vanishing correlation at two particular values of $\gamma$, and this signifies a \textit{first}-order phase transition. For $\mathfrak{c}_j^{NEM}\rightarrow 0$, using Eqs.~(\ref{eq2.25}, \ref{eq2.26}, \ref{eq2.27}), we get the condition: $\sinh^2{\beta\mathcal{N}} = 0$, i.e., $\mathcal{N} = \frac{1}{2}(\mathbb{B}\sin^2{\gamma}-\mathbb{C}\cos^2{\gamma}) = 0$, and this gives us the same expression for the critical $\gamma$'s as before, i.e., $\gamma_{A, B} = \tan^{-1}\sqrt{\mathbb{C}/\mathbb{B}}$. For $\mathbb{B} = 5.0$ and $\mathbb{C} = 2.0$, we have $\gamma = \tan^{-1}\sqrt{{2}/{5}} = 0.5639~\text{or}~2.5777$. Therefore, we have $\gamma_A = 0.5639$ and $\gamma_B = 2.5777$ for the given values of $\mathbb{B} = 5.0$ and $\mathbb{C} = 2.0$ (see, Figs.~\ref{fig3a}, \ref{fig3b}) and they are independent of the \textit{noise} $\beta$. 

When $\beta$ is finite, as $\gamma \in [0, \gamma_A]$ gradually increases, we find an initial drop in the strategy correlation, reaching a minimum at $\gamma = \gamma_A$. This is mainly attributed to the fact that while $\gamma \rightarrow \gamma_1$, the quantum players tend to shift from classical \textit{defect} to \textit{quantum}, resulting in a lowering of correlation among the quantum players. Then, as we increase the value of $\gamma$ from $\gamma_A$ to $\gamma = \frac{\pi}{2}$, the correlation increases since most of the quantum players now play \textit{quantum} as $\gamma \rightarrow \frac{\pi}{2}$. At maximal entanglement, i.e., $\gamma = \frac{\pi}{2}$, the correlation is maximum since, in this case, all the quantum players play \textit{quantum}, and they do not shift to \textit{defect} on minute changes in $\gamma$. However, as we further increase the $\gamma$ value from $\frac{\pi}{2}$ to $\gamma_B$, the strategy correlation again decreases due to a strategy shift from \textit{quantum} to \textit{defect} among the quantum players, reaching a minimum at $\gamma_B$. When $\gamma>\gamma_B$, the majority of quantum players switch to \textit{defect}, thus increasing the correlation. When $\gamma=0~\text{or}~\pi$, a large fraction of quantum players play \textit{defect}. In the Z-N limit, i.e., $T\rightarrow 0$ (or, $\beta \rightarrow \infty$), $\mathfrak{c}_j^{NEM} \rightarrow 1,~\forall~\gamma\neq \{\gamma_A, \gamma_B\}$. Here, we also observe two sharp discontinuous phase transition peaks, at $\gamma_A$ and $\gamma_B$ (see, Fig.~\ref{fig3b}), when all quantum players shift from \textit{defect} to \textit{quantum} (at $\gamma=\gamma_A$) and back to \textit{defect} (at $\gamma=\gamma_B$) as the $\gamma$ value increases from $0$ to $\pi$. In the I-N limit, i.e., $T\rightarrow \infty$ (or, $\beta \rightarrow 0$), $\mathfrak{c}_j^{NEM} \rightarrow 0,~\forall~\gamma$, implying that the quantum players opt for their strategies randomly, resulting in vanishing correlation.

\subsubsection{\underline{ABM}}
For the given values of $\mathcal{T} = 0$ and $\mathcal{F} = \frac{(\mathbb{B}\sin^2{\gamma}-\mathbb{C}\cos^2{\gamma})}{2}$ in Eq.~(\ref{eq2.20}), we have the Energy matrix $\Delta = -{\Lambda}$ (see, Eq.~(\ref{eq2.11})). Thus,
\begin{equation}
    \Delta = \left[\begin{array}{c c}  
         -(\mathbb{B-C}) & -( \mathbb{B}\sin^2{\gamma}-\mathbb{C}\cos^2{\gamma})\\
         -(\mathbb{B}\cos^2{\gamma} - \mathbb{C}\sin^2{\gamma}) & 0
    \end{array}\right].
        \label{eq3.4}
\end{equation}
Following the algorithm given in Sec.~\ref{ABM}, for $\mathbb{B} = 5.0$ and $\mathbb{C} = 2.0$, we determine the ABM correlation ($\mathfrak{c}_{j}^{ABM}$), and its variation with the entanglement $\gamma$ is shown in Figs.~\ref{fig3a}, \ref{fig3b}. We again observe exactly the same results as obtained for correlation, via NEM, in the finite and limiting values of $\beta$. When $\beta$ is finite, as $\gamma \in [0, \gamma_A]$ gradually increases, we find an initial drop in the strategy correlation, reaching a minimum at $\gamma = \gamma_A$. Then, as we increase the value of $\gamma$ from $\gamma_A$ to $\gamma = \frac{\pi}{2}$, the correlation increases. At maximal entanglement, i.e., $\gamma = \frac{\pi}{2}$, the correlation is maximum. As we further increase the $\gamma$ value from $\frac{\pi}{2}$ to $\gamma_B$, the correlation again decreases, reaching a minimum at $\gamma_B$.

In the Z-N limit, i.e., $T\rightarrow 0$ (or, $\beta \rightarrow \infty$), $\mathfrak{c}_j^{ABM} \rightarrow 1,~\forall~\gamma\neq \{\gamma_A, \gamma_B\}$. As shown in Fig.~\ref{fig3b}, we observe two sharp phase transition peaks, at $\gamma_A$ and $\gamma_B$, when all quantum players shift from \textit{defect} to \textit{quantum} (at $\gamma=\gamma_A$) and back to \textit{defect} (at $\gamma=\gamma_B$) as the $\gamma$ value increases from $0$ to $\pi$. In the I-N limit, i.e., $T\rightarrow \infty$ (or, $\beta \rightarrow 0$), $\mathfrak{c}_j^{ABM} \rightarrow 0,~\forall~\gamma$, implying that the quantum players opt for their strategies randomly, resulting in vanishing correlation.

\subsubsection{\underline{Analysis of correlation}}
The analysis of correlation: $\mathfrak{c}_j^{NEM}$ as well as $\mathfrak{c}_j^{ABM}$, will be done in this subsection. From Fig.~\ref{fig:3} we observe that for increasing values of $\beta$, for both \textit{minimal} (i.e., $\gamma\rightarrow0~\text{or}~\pi$) and \textit{maximal entanglement} (i.e., $\gamma\rightarrow\frac{\pi}{2}$), $\mathfrak{}c_j^{ABM} = \mathfrak{c}_j^{NEM} \rightarrow 1$, indicating maximum correlation among the strategies adopted by the quantum players. In both $\gamma\rightarrow 0$ and $\gamma\rightarrow\pi$ limits, for increasing $\beta$, a large fraction of quantum players play the \textit{defect} strategy, leading to maximum correlation. Similarly, in the $\gamma\rightarrow \frac{\pi}{2}$ limit, almost all quantum players adopt the \textit{quantum} strategy, and this also leads to a maximum correlation. In the Z-N (or, $\beta\rightarrow\infty$) limit, we observe two \textit{first}-order phase transition peaks, as shown in Fig.~\ref{fig3b}, at the two critical values of $\gamma$ (i.e., at $\gamma_A$ and $\gamma_B$). The values of $\gamma_{A, B}$, where we observe the phase transitions, depend on the payoffs $\mathbb{B}$ and $\mathbb{C}$ via the relation: $\gamma_{A, B} = \tan^{-1}\sqrt{\mathbb{C}/\mathbb{B}}$, and both $\mathbb{B}$ as well as $\mathbb{C}$ can also induce phase transition(s) in a similar way to what we discussed in the case of game magnetization $\mu$. 
\begin{figure*}[!ht]
    \centering
    \begin{subfigure}[b]{0.95\columnwidth}
        \centering
        \includegraphics[width = \textwidth]{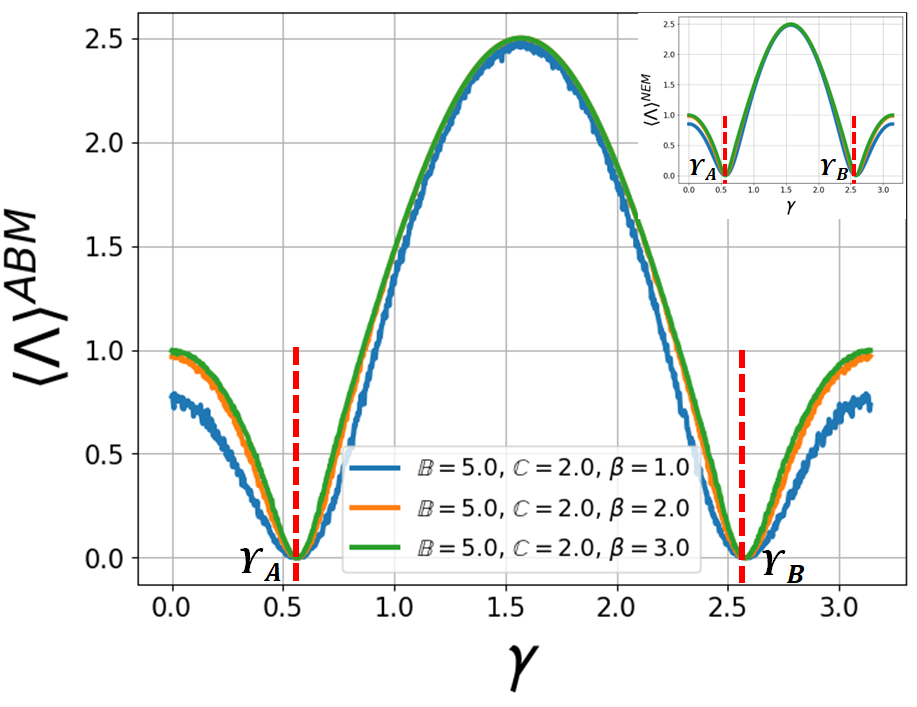}
        \caption{$\langle\Lambda\rangle^{ABM/NEM}$ vs $\gamma$}
        \label{fig4a}
    \end{subfigure}
    \begin{subfigure}[b]{0.95\columnwidth}
        \centering
        \includegraphics[width = \textwidth]{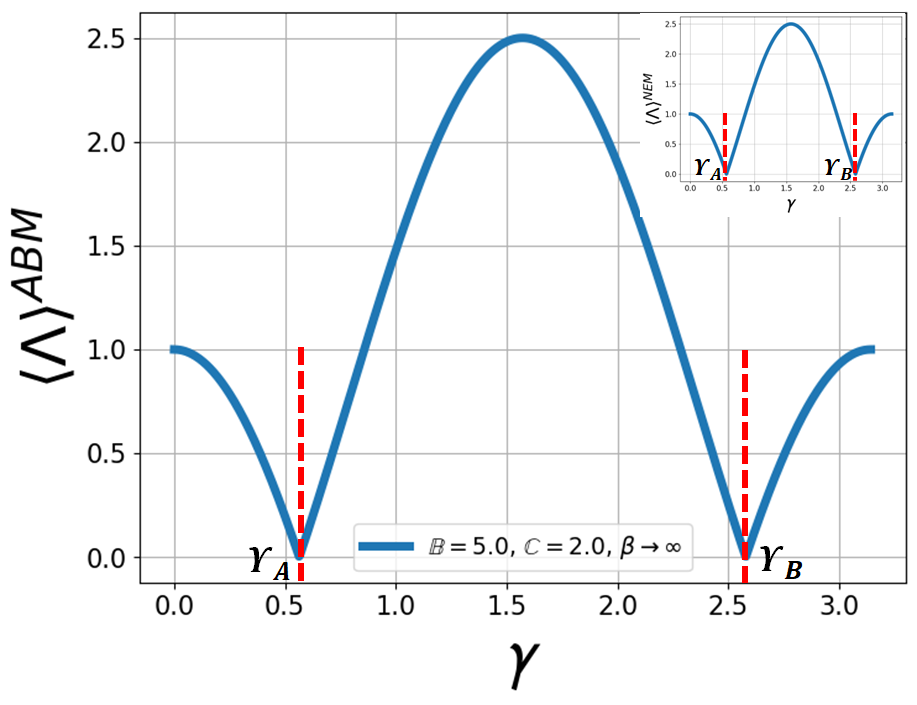}
        \caption{$\langle\Lambda\rangle^{ABM/NEM}$ vs $\gamma$ in $\beta\rightarrow \infty$ (or, $T\rightarrow 0$) limit}
        \label{fig4b}
    \end{subfigure}
    \caption{\centering{ABM and NEM (in \textit{insets}): \textbf{Player's payoff average} $\langle\Lambda\rangle$ vs $\gamma$ for (a) $ \beta= 1, 2, 3$and (b)$\beta\rightarrow \infty$. Other parameters common to both are \textbf{reward} $\mathbb{R}= (\mathbb{B-C}) =3.0$, \textbf{sucker's payoff} $\mathbb{S}= \mathbb{-C}= -2.0$, \textbf{temptation} $\mathbb{T}= \mathbb{B}=5.0$, \textbf{punishment} $\mathbb{P}=0.0$, $\gamma_A = 0.5639$ and $\gamma_B=2.5777$ in QuPD.}}
    \label{fig:4}
\end{figure*}

\subsection{Player's payoff average $(\langle\Lambda\rangle)$}
\subsubsection{\underline{NEM}}
Using the given values of $\mathbb{B} = 5.0$, $\mathbb{C} = 2.0$ and the expression of $\langle\Lambda\rangle^{NEM}$ in Eq.~(\ref{eq2.28}), we plot the variation of $\langle\Lambda\rangle^{NEM}$ with regard to changing values of $\gamma$ and they are given in the \textit{insets} of Figs.~\ref{fig4a}, \ref{fig4b}. In both Figs.~\ref{fig4a} and \ref{fig4b}, we observe a vanishing payoff average at two particular values of $\gamma$, signifying a \textit{first}-order phase transition. For $\langle\Lambda\rangle^{NEM}\rightarrow 0$, using Eq.~(\ref{eq2.28}), we get the condition: $\mathcal{N} = \frac{1}{2}(\mathbb{B}\sin^2{\gamma}-\mathbb{C}\cos^2{\gamma}) = 0$, and this gives us the same expression for the critical $\gamma$'s as seen for the previous indicators, i.e., $\gamma_{A, B} = \tan^{-1}\sqrt{\mathbb{C}/\mathbb{B}}$. For $\mathbb{B} = 5.0$ and $\mathbb{C} = 2.0$, we have $\gamma = \tan^{-1}\sqrt{{2}/{5}} = 0.5639~\text{or}~2.5777$. Therefore, we have $\gamma_A = 0.5639$ and $\gamma_B = 2.5777$ for given values of $\mathbb{B} = 5.0$ and $\mathbb{C} = 2.0$ (see, Figs.~\ref{fig4a}, \ref{fig4b}) and they are independent of the \textit{noise} $\beta$.

When $\beta$ is finite, as $\gamma \in [0, \gamma_A]$ gradually increases, we find an initial drop in the $\langle\Lambda\rangle^{NEM}$ value, reaching a minimum at $\gamma = \gamma_A$. This is mainly attributed to the fact that while $\gamma$ approaches $\gamma_1$, the quantum players tend to switch from classical \textit{defect} to \textit{quantum}, resulting in a lowering of feasible payoffs accessible to all quantum players, see the payoff matrix given in Eq.~(\ref{eq2.11}) for the payoffs associated with $[\mathbb{Q}, \mathfrak{D}]$ or $[\mathfrak{D}, \mathbb{Q}]$ case. Then, as we increase the value of $\gamma$ from $\gamma_A$ to $\gamma = \frac{\pi}{2}$, $\langle\Lambda\rangle^{NEM}$ increases since most of the quantum players have switched to \textit{quantum} as $\gamma$ approaches $\frac{\pi}{2}$, resulting in a comparatively better average payoff than the classical strategy (\textit{defect}) payoffs (see, [$\mathbb{Q, Q}$]-payoff in Eq.~(\ref{eq2.11})). At maximal entanglement, i.e., $\gamma = \frac{\pi}{2}$, $\langle\Lambda\rangle^{NEM}$ is maximum since, in this case, all the quantum players play \textit{quantum} and they do not shift to \textit{defect} on minute changes in $\gamma$, resulting in the best possible payoff available to each quantum player. However, as we further increase the $\gamma$ value from $\frac{\pi}{2}$ to $\gamma_B$, $\langle\Lambda\rangle^{NEM}$ again decreases due to a switch from \textit{quantum} to \textit{defect} among the quantum players, reaching a minimum at $\gamma_B$. When $\gamma>\gamma_B$, the majority of quantum players switch over to \textit{defect}, thus increasing the classical average payoff associated with the \textit{defect} strategy. When $\gamma=0~\text{or}~\pi$, a large fraction of quantum players play \textit{defect}.

In the Z-N limit, i.e., $T\rightarrow 0$ (or, $\beta \rightarrow \infty$), $\langle\Lambda\rangle^{NEM} \rightarrow \frac{|\mathbb{B}\sin^2\gamma - \mathbb{C}\cos^2\gamma|}{2},~\forall~\gamma$. This shows that the player's payoff average depends on the entanglement $\gamma$ and always satisfies $\langle\Lambda\rangle^{NEM}\geq 0$, irrespective of the payoffs. In the I-N limit, i.e., $T\rightarrow \infty$ (or, $\beta \rightarrow 0$), $\langle\Lambda\rangle^{NEM} \rightarrow 0,~\forall~\gamma$, implying that the quantum players opt for their strategies randomly, leading to payoff randomization. By summing up the four payoffs (or, the matrix elements) given in the energy (or, $-ve$ payoff) matrix in Eq.~(\ref{eq3.6}), we indeed get a vanishing average payoff per quantum player.

\subsubsection{\underline{ABM}}
To determine the individual player's payoff average, we start with the given values of $\mathcal{T} = 0$ and $\mathcal{F} = \frac{(\mathbb{B}\sin^2{\gamma}-\mathbb{C}\cos^2{\gamma})}{2}$ in Eq.~(\ref{eq2.20}), respectively. The Energy matrix $\Delta = -{\Lambda}$ (see, Eq.~(\ref{eq2.11})) is again given as,
\begin{equation}
    \Delta = \left[\begin{array}{c c}  
         -(\mathbb{B-C}) & -( \mathbb{B}\sin^2{\gamma}-\mathbb{C}\cos^2{\gamma})\\
         -(\mathbb{B}\cos^2{\gamma} - \mathbb{C}\sin^2{\gamma}) & 0
    \end{array}\right].
        \label{eq3.5}
\end{equation}
However, this time, instead of considering the $\Delta$ in Eq.~(\ref{eq3.5}), we consider a modified $\Delta$ (say, $\Delta'$)\cite{ref12} whose elements are the linear transformations of the original energy matrix elements given in Eq.~(\ref{eq3.5}) (see, the set of linear transformations given in Eq.~(\ref{eq2.16})). Both these matrices have a one-to-one correspondence, and hence, the Nash equilibrium is preserved. So, we redefine the energy matrix as,
\begin{equation}
    \Delta' = \left[\begin{array}{c c}  
         -\frac{(\mathbb{B}\sin^2{\gamma}-\mathbb{C}\cos^2{\gamma})}{2} & -\frac{(\mathbb{B}\sin^2{\gamma}-\mathbb{C}\cos^2{\gamma})}{2}\\
         \frac{(\mathbb{B}\sin^2{\gamma}-\mathbb{C}\cos^2{\gamma})}{2} & \frac{(\mathbb{B}\sin^2{\gamma}-\mathbb{C}\cos^2{\gamma})}{2}
    \end{array}\right].
        \label{eq3.6}
\end{equation}
\begin{figure*}[!ht]
    \centering
    \begin{subfigure}[b]{0.95\columnwidth}
        \centering
        \includegraphics[width = \textwidth]{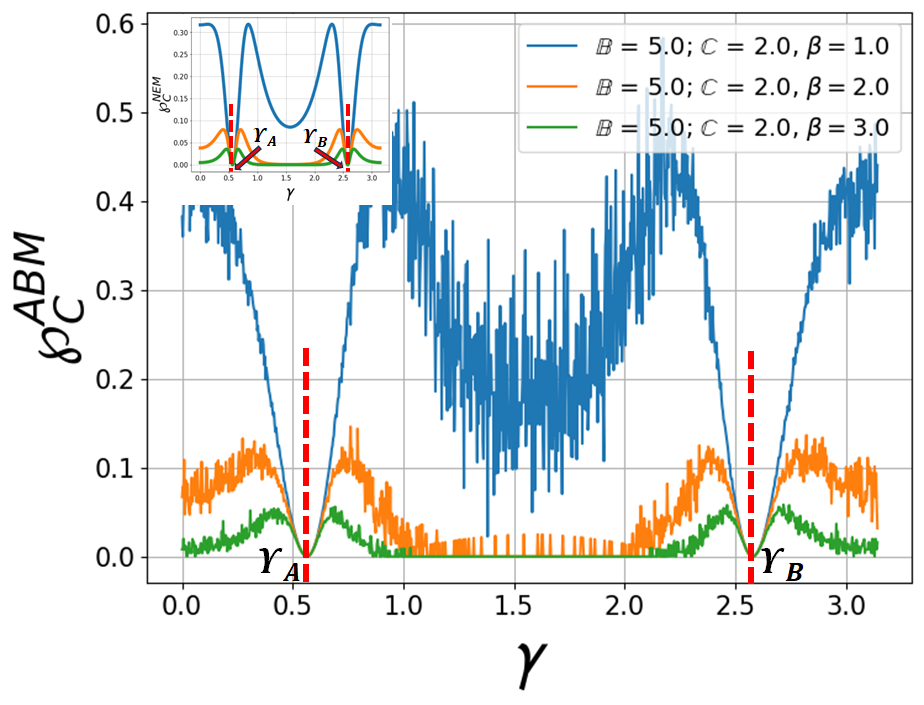}
        \caption{$\wp_{C}^{ABM/NEM}$ vs $\gamma$}
        \label{fig5a}
    \end{subfigure}
    \begin{subfigure}[b]{0.95\columnwidth}
        \centering
        \includegraphics[width = \textwidth]{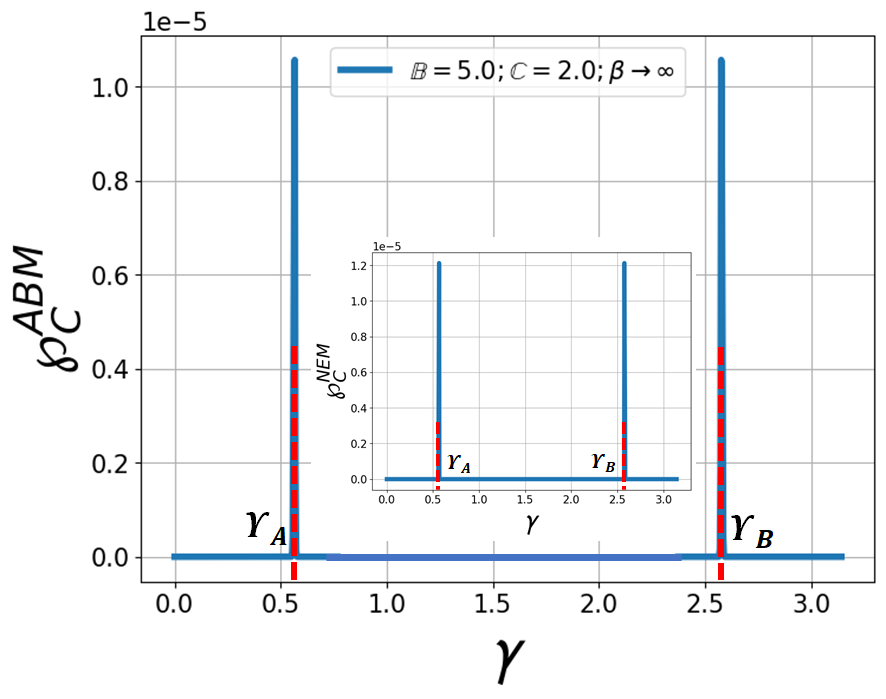}
        \caption{$\wp_{C}^{ABM/NEM}$ vs $\gamma$ in $\beta\rightarrow \infty$ (or, $T\rightarrow 0$) limit}
        \label{fig5b}
    \end{subfigure}
    \caption{\centering{ABM and NEM (in \textit{insets}): \textbf{Payoff capacity} $\wp_{C}$ vs \textbf{entanglement} $\gamma$ for (a) $\beta=1, 2, 3$ and (b) $\beta\rightarrow \infty$. Other parameters common to both are \textbf{reward} $\mathbb{R}=3.0$, \textbf{sucker's payoff} $\mathbb{S}=0.0$, \textbf{temptation} $\mathbb{T}=5.0$, \textbf{punishment} $\mathbb{P}=1.0$, $\gamma_A = 0.5639$ and $\gamma_B=2.5777$ in QuPD.}}
    \label{fig:5}
\end{figure*}
Both $\Delta$ in Eq.~(\ref{eq3.5}) and $\Delta'$ in Eq.~(\ref{eq3.6}) are equivalent to each other. Now, by following the algorithm given in Sec.~\ref{ABM}, for $\mathbb{B} = 5.0$ and $\mathbb{C} = 2.0$, we determine the ABM player's payoff average ($\langle\Lambda\rangle^{ABM}$), and its variation with the entanglement $\gamma$ is shown in Figs.~\ref{fig4a}, \ref{fig4b}. We again observe exactly the same results as obtained for $\langle\Lambda\rangle$, via NEM, in the finite and limiting values of $\beta$.

When $\beta$ is finite, as $\gamma \in [0, \gamma_A]$ gradually increases, we find an initial drop in the $\langle\Lambda\rangle^{ABM}$ value, reaching a minimum at $\gamma = \gamma_A$. Then, as we increase the value of $\gamma$ from $\gamma_A$ to $\gamma = \frac{\pi}{2}$, $\langle\Lambda\rangle^{ABM}$ increases since most of the quantum players switch to \textit{quantum} as $\gamma$ approaches $\frac{\pi}{2}$. At maximal entanglement, i.e., $\gamma = \frac{\pi}{2}$, $\langle\Lambda\rangle^{ABM}$ is maximum since, in this case, all the quantum players play \textit{quantum}, and they do not shift to \textit{defect} on minute changes in $\gamma$. However, as we further increase the $\gamma$ value from $\frac{\pi}{2}$ to $\gamma_B$, $\langle\Lambda\rangle^{ABM}$ again decreases due to a strategy shift from \textit{quantum} to \textit{defect} among the quantum players, reaching a minimum at $\gamma_B$. When $\gamma>\gamma_B$, the majority of quantum players choose to switch over to \textit{defect}. In the Z-N limit, i.e., $T\rightarrow 0$ (or, $\beta \rightarrow \infty$), $\langle\Lambda\rangle^{ABM} \rightarrow \frac{|\mathbb{B}\sin^2\gamma - \mathbb{C}\cos^2\gamma|}{2},~\forall~\gamma$. This shows that the player's payoff average, similar to what we observed in ABM, depends on the entanglement $\gamma$ and always satisfies $\langle\Lambda\rangle^{ABM}\geq 0$, irrespective of the payoffs. In the I-N limit, i.e., $T\rightarrow \infty$ (or, $\beta \rightarrow 0$), $\langle\Lambda\rangle^{ABM} \rightarrow 0,~\forall~\gamma$, implying that the quantum players opt for their strategies randomly, leading to payoff randomization.

\subsubsection{\label{ipap}\underline{Analysis of Player's payoff average}}
The analysis of the Player's payoff average: $\langle\Lambda\rangle^{NEM}$ as well as $\langle\Lambda\rangle^{ABM}$, will be done in this subsection. From Fig.~\ref{fig:4}, we observe that for increasing values of $\beta$, in \textit{minimal entanglement} (i.e., $\gamma\rightarrow 0~\text{or}~\pi$), $\langle\Lambda\rangle^{ABM} = \langle\Lambda\rangle^{NEM} \rightarrow \frac{\mathbb{C}}{2} = 1$, for given $\mathbb{C}=2.0$, indicating the average payoff associated with the classical \textit{defect} ($\mathfrak{D}$) strategy, which in fact is the dominant strategy in this case. Meanwhile, in \textit{maximal entanglement} (i.e., $\gamma\rightarrow \frac{\pi}{2}$), when all quantum players play \textit{quantum}, $\langle\Lambda\rangle^{ABM} = \langle\Lambda\rangle^{NEM} \rightarrow \frac{\mathbb{B}}{2} = \frac{5}{2}$, for given $\mathbb{B}=5.0$, indicating the average payoff associated with the \textit{quantum} ($\mathbb{Q}$) strategy. For any $(\mathbb{B,C})$ that satisfies the criteria: $\mathbb{B}> \mathbb{C}>0$, the average payoff associated with the \textit{quantum} strategy always exceeds the average payoff associated with the \textit{defect} strategy and this results in a large fraction of quantum players switching their strategies from $\mathfrak{D}\rightarrow\mathbb{Q}$ when the entanglement $\gamma \in [\gamma_A, \gamma_B]$. For finite as well as limiting values of $\beta$, we observe a vanishing average payoff at both the critical $\gamma$ points, i.e., $\gamma_A$ and $\gamma_B$, and this signifies the change in phases from \textit{defect} to \textit{quantum} and vice-versa at $\gamma_A$ and $\gamma_B$, respectively. Here too, the values of $\gamma_{A, B}$, where we observe the phase transitions, depend on the payoffs $\mathbb{B}$ and $\mathbb{C}$ via the relation: $\gamma_{A, B} = \tan^{-1}\sqrt{\mathbb{C}/\mathbb{B}}$, and both $\mathbb{B}$ as well as $\mathbb{C}$ can induce phase transition(s).

\subsection{Payoff capacity $(\wp_C)$}

\subsubsection{\underline{NEM}}
Using the given values of $\mathbb{B} = 5.0$, $\mathbb{C} = 2.0$ and the expression of $\wp_C^{NEM}$ in Eq.~(\ref{eq2.29}), we plot the variation of $\wp_C^{NEM}$ with regard to changing values of $\gamma$ and they are shown in the \textit{insets} of Figs.~\ref{fig5a}, \ref{fig5b}. In the $\beta\rightarrow\infty$ (i.e., Z-N) limit, we observe \textit{two} sharp discontinuous peaks at two values of $\wp_C^{NEM}$, indicating \textit{first}-order phase transitions. To determine these critical $\gamma$ values where $\wp_C^{NEM}\rightarrow\infty$, in the $\beta\rightarrow\infty$ limit, we equate $\frac{1}{\wp_C^{NEM}}\rightarrow 0$ (see, Eq.~(\ref{eq2.29})) and we again get the condition: $\frac{1}{2}(\mathbb{B}\sin^2{\gamma}-\mathbb{C}\cos^2{\gamma}) = 0$, and this gives us the same expression for the critical $\gamma$'s, i.e., $\gamma_{A, B} = \tan^{-1}\sqrt{\mathbb{C}/\mathbb{B}}$. For $\mathbb{B} = 5.0$ and $\mathbb{C} = 2.0$, we have $\gamma = \tan^{-1}\sqrt{{2}/{5}} = 0.5639~\text{or}~2.5777$. Therefore, we have $\gamma_A = 0.5639$ and $\gamma_B = 2.5777$ for the given values of $\mathbb{B} = 5.0$, $\mathbb{C} = 2.0$ (see, Figs.~\ref{fig5a}, \ref{fig5b}) and they are independent of the \textit{noise} $\beta$. 

When $\beta$ is finite, as $\gamma \in [0, \gamma_A]$ gradually increases, we find an initial drop in the $\wp_C^{NEM}$ value, reaching a minimum at $\gamma = \gamma_A$. This is mainly attributed to the fact that while $\gamma$ approaches $\gamma_1$, the quantum players tend to switch from \textit{defect} to \textit{quantum}, resulting in a lowering of the average feasible payoffs accessible to the quantum players, which further leads to a minimal change in the average payoff when there is a unit change in the \textit{noise}. Then, as we increase the value of $\gamma$ from $\gamma_A$ to $\gamma = \frac{\pi}{2}$, $\wp_C^{NEM}$ first increases and then again it decreases since within this range of $\gamma$, the quantum players update their strategies from \textit{defect} to \textit{quantum}, and hence we initially observe a significant alteration in the payoff owing to a unit change in the noise. However, when all quantum players play \textit{quantum}, then the payoffs do not change significantly with regards to a unit change in noise, resulting in a decrease of $\wp_C^{NEM}$ as $\gamma$ approaches $\frac{\pi}{2}$. However, as we further increase the $\gamma$ value from $\frac{\pi}{2}$ to $\gamma_B$, $\wp_C^{NEM}$ value again initially increases and then decreases due to the same logic as mentioned before, but now the strategy shift gets reversed, and the quantum players tend to shift from \textit{quantum} to \textit{defect}. When $\gamma=0~\text{or}~\pi$, a large fraction of quantum players play \textit{defect}.
\begin{table*}[!ht]
\centering
\renewcommand{\arraystretch}{1.6}
\resizebox{1.3\columnwidth}{!}{%
\begin{tabular}{|cc|c|c|c|}
\hline
\multicolumn{2}{|c|}{\textit{\textbf{For QuPD game}}} & \multicolumn{1}{c|}{\textbf{ABM}} & \multicolumn{1}{c|}{\textbf{NEM}} \\ \hline
\multicolumn{1}{|c|}{\multirow{2}{*}{\Large$\mathbf{\mu}$}} & \multicolumn{1}{c|}{$\beta \rightarrow 0$} & \multicolumn{1}{c|}{0, $\forall~\gamma$} & \multicolumn{1}{c|}{0, $\forall~\gamma$} \\ \cline{2-4} 
\multicolumn{1}{|c|}{} & \multicolumn{1}{c|}{$\beta \rightarrow \infty$} & \begin{tabular}[c]{@{}c@{}}$+ 1$, $\forall~\gamma_A < \gamma < \gamma_B$\\ $-1$, $\forall~\gamma<\gamma_A$ or $\gamma>\gamma_B$\end{tabular} & \begin{tabular}[c]{@{}c@{}}$+ 1$, $\forall~\gamma_A < \gamma < \gamma_B$\\ $-1$, $\forall~\gamma<\gamma_A$ or $\gamma>\gamma_B$\end{tabular} \\ \hline
\multicolumn{1}{|c|}{\multirow{2}{*}{\Large$\mathbf{\chi_\mathfrak{\gamma}}$}} & \multicolumn{1}{c|}{$\beta \rightarrow 0$} & \multicolumn{1}{c|}{$\frac{(\mathbb{B+C})}{2} \sin{(2\gamma)}, \forall~\gamma$} & \multicolumn{1}{c|}{$\frac{(\mathbb{B+C})}{2} \sin{(2\gamma)}, \forall~\gamma$} \\ \cline{2-4} 
\multicolumn{1}{|c|}{} & \multicolumn{1}{c|}{$\beta \rightarrow \infty$} & \multicolumn{1}{c|}{0, $\forall~\gamma\neq \{\gamma_A, \gamma_B\}$} & \multicolumn{1}{c|}{0, $\forall~\gamma\neq \{\gamma_A, \gamma_B\}$} \\ \hline
\multicolumn{1}{|l|}{\multirow{2}{*}{\Large$\mathbf{\mathfrak{c}_j}$}} & $\beta \rightarrow 0$ & $0,~\forall~\gamma$ & $0,~\forall~\gamma$ \\ \cline{2-4} 
\multicolumn{1}{|l|}{} & $\beta \rightarrow \infty$ & \multicolumn{1}{c|}{1, $\forall~\gamma\neq \{\gamma_A, \gamma_B\}$} & \multicolumn{1}{c|}{1, $\forall~\gamma\neq \{\gamma_A, \gamma_B\}$} \\ \hline
\multicolumn{1}{|l|}{\multirow{2}{*}{\Large$\mathbf{\Lambda}$}} & $\beta \rightarrow 0$ & $0,~\forall~\gamma$ & $0,~\forall~\gamma$ \\ \cline{2-4} 
\multicolumn{1}{|l|}{} & $\beta \rightarrow \infty$ & $\frac{|\mathbb{B}\sin^2{\gamma} - \mathbb{C}\cos^2{\gamma}|}{2},~\forall~\gamma$ & $\frac{|\mathbb{B}\sin^2{\gamma} - \mathbb{C}\cos^2{\gamma}|}{2},~\forall~\gamma$ \\ \hline
\multicolumn{1}{|l|}{\multirow{2}{*}{\Large$\mathbf{\wp_c}$}} & $\beta \rightarrow 0$ & $\frac{(\mathbb{B}\sin^2{\gamma} - \mathbb{C}\cos^2{\gamma})^2}{3},~\forall~\gamma$ & $\frac{(\mathbb{B}\sin^2{\gamma} - \mathbb{C}\cos^2{\gamma})^2}{3},~\forall~\gamma$ \\ \cline{2-4} 
\multicolumn{1}{|l|}{} & $\beta \rightarrow \infty$ & $0,~\forall~\gamma\neq \{\gamma_A, \gamma_B\}$ & $0,~\forall~\gamma\neq \{\gamma_A, \gamma_B\}$ \\ \hline
\end{tabular}%
}
\caption{QuPD game with  reward $\mathbb{R}= (\mathbb{B-C}) = 3.0$, \textbf{sucker's payoff} $\mathbb{S}= \mathbb{-C}=-2.0$, \textbf{temptation} $\mathbb{T} = \mathbb{B} =5.0$, \textbf{punishment} $\mathbb{P}=0.0$, inter-site distance j, $\gamma_A = 0.5639$, $\gamma_B=2.5777$ and measure of noise $\beta$.
}
\label{tab:table1}
\end{table*}

In the Z-N limit, i.e., $T\rightarrow 0$ (or, $\beta \rightarrow \infty$), $\wp_C^{NEM} \rightarrow 0,~\forall~\gamma\neq \{\gamma_A, \gamma_B\}$, indicating two phase transition peaks at $\gamma_A$ and $\gamma_B$. In the Z-N limit, we see no variation in a player's payoff (i.e., the payoff becomes \textit{constant}), which leads to a vanishing payoff capacity except at the two critical values of $\gamma_A$ and $\gamma_B$, where we get vanishing payoffs (see, Sec.~\ref{ipap}) and this indicates the change in phases, i.e., from $\mathfrak{D}\rightarrow\mathbb{Q}$ and vice-versa, resulting in a small yet noticeable peak in the payoff capacity plot (see, Fig.~\ref{fig5b}). In the I-N limit, i.e., $T\rightarrow \infty$ (or, $\beta \rightarrow 0$), $\wp_C^{NEM} \rightarrow \frac{(\mathbb{B}\sin^2\gamma - \mathbb{C}\cos^2\gamma)^2}{3},~\forall~\gamma$, implying that the quantum players opt for their strategies randomly. For finite non-zero $\beta$, $\wp_C^{NEM}$ is always $+ve$, indicating that $\langle\Lambda\rangle^{NEM}$ changes at a faster rate with increasing \textit{noise}.

\subsubsection{\underline{ABM}}
Similar to $\langle\Lambda\rangle^{ABM}$, to determine the payoff capacity $\wp_C^{ABM}$, we start with the given values of $\mathcal{T} = 0$ and $\mathcal{F} = \frac{(\mathbb{B}\sin^2{\gamma}-\mathbb{C}\cos^2{\gamma})}{2}$ in Eq.~(\ref{eq2.20}), respectively. Similar to the previous case, here also, we consider the modified energy matrix whose elements are the linear transformations of the original energy matrix elements given in Eq.~(\ref{eq3.5}) (see, the set of linear transformations given in Eq.~(\ref{eq2.16})). Both these matrices have a one-to-one correspondence, and hence, the Nash equilibrium is preserved. So, we have the modified energy matrix as,
\begin{equation}
    \Delta' = \left[\begin{array}{c c}  
         -\frac{(\mathbb{B}\sin^2{\gamma}-\mathbb{C}\cos^2{\gamma})}{2} & -\frac{(\mathbb{B}\sin^2{\gamma}-\mathbb{C}\cos^2{\gamma})}{2}\\
         \frac{(\mathbb{B}\sin^2{\gamma}-\mathbb{C}\cos^2{\gamma})}{2} & \frac{(\mathbb{B}\sin^2{\gamma}-\mathbb{C}\cos^2{\gamma})}{2}
    \end{array}\right].
        \label{eq3.7}
\end{equation}
Both $\Delta$ in Eq.~(\ref{eq3.5}) and $\Delta'$ in Eq.~(\ref{eq3.7}) are equivalent to each other. Now, by following the algorithm given in Sec.~\ref{ABM}, for $\mathbb{B} = 5.0$ and $\mathbb{C} = 2.0$, we determine the ABM payoff capacity ($\wp_C^{ABM}$), and its variation with the entanglement $\gamma$ is shown in Figs.~\ref{fig5a}, \ref{fig5b}. We again observe exactly the same results obtained for $\wp_C^{ABM}$ and $\wp_C^{NEM}$ in the finite and limiting values of $\beta$.

\subsubsection{\underline{Analysis for payoff capacity}}
The analysis of the payoff capacity: $\wp_C^{NEM}$ as well as $\wp_C^{ABM}$, will be done in this subsection. From Fig.~\ref{fig:5}, for increasing values of $\beta$, we observe that for all $\gamma \rightarrow \{0,~\frac{\pi}{2},~\pi\}$, i.e., for both \textit{minimal} (i.e., $\gamma\rightarrow 0~\text{or}~\pi$) and \textit{maximal entanglement} (i.e., $\gamma\rightarrow\frac{\pi}{2}$), $\wp_C^{ABM} = \wp_C^{NEM} \rightarrow 0$, indicating no phase transition among the quantum players for a unit change in \textit{noise}. This can also be verified from the player's payoff average result (see, Fig.~\ref{fig:4}) where we see that for finite as well as limiting values of $\beta$, when $\gamma \rightarrow \{0,~\frac{\pi}{2},~\pi\}$, a large fraction of quantum players choose either \textit{defect} (for $\gamma\rightarrow0$ or $\pi$) or \textit{quantum} (for $\gamma\rightarrow\frac{\pi}{2}$), and this leads to a vanishing $\wp_C$. Interestingly, in the Z-N (or, $\beta\rightarrow\infty$) limit, we observe two \textit{first}-order phase transition peaks, as shown in Fig.~\ref{fig5b}, at the two critical values of $\gamma$ (i.e., at $\gamma_A$ and $\gamma_B$) and this signifies the change in phases from \textit{defect} to \textit{quantum} and vice-versa at $\gamma_A$ and $\gamma_B$, respectively. The values of $\gamma_{A, B}$, where we observe the phase transitions, depend on the payoffs $\mathbb{B}$ and $\mathbb{C}$ via the relation: $\gamma_{A, B} = \tan^{-1}\sqrt{\mathbb{C}/\mathbb{B}}$, and both $\mathbb{B}$ as well as $\mathbb{C}$ can induce phase transition(s).

\section{\label{conc}Conclusion}
In this tutorial, we sought to understand the emergence of cooperative behaviour among an infinite number of quantum players playing the quantum Prisoner's dilemma (QuPD) game by comparing a numerical technique, i.e., Agent-based modelling (ABM), with the analytical NEM method. In the \textit{TL} of the \textit{one-shot} game setup, we studied five different indicators, i.e., game magnetization ($\mu$), entanglement susceptibility ($\chi_{\gamma}$), correlation ($\mathfrak{c}_j$), player's payoff average ($\langle\Lambda\rangle$) and payoff capacity ($\wp_C$), to understand the phase transitions occurring in QuPD. Table-\ref{tab:table1} summarises the outcomes for each of the five indicators in the limiting $\beta$ cases. For all the five indicators in question, we observed that \textit{quantum} ($\mathbb{Q}$) remains the dominant strategy for a large fraction of quantum players within a particular entanglement ($\gamma$) range (i.e., $\gamma\in [\gamma_A, \gamma_B]$). Within this $\gamma$-range, $\mathbb{Q}$ becomes the \textit{Nash equilibrium} strategy as well as the \textit{Pareto optimal} strategy for the quantum players. For the payoff values $\mathbb{B}=5.0$ and $\mathbb{C}=2.0$, we found critical $\gamma$ values: $\gamma_A = 0.5639$ and $\gamma_B = 2.5777$, within which all quantum players play $\mathbb{Q}$. For other values of $\gamma$, i.e., $\gamma\in [0,\gamma_A)\cup(\gamma_B, \pi]$, a large fraction of quantum players play the classical \textit{defect} ($\mathcal{D}$). For the maximally entangled case, i.e., $\gamma=\pi/2$, we observed that the player's payoff average (corresponding to all quantum players playing $\mathbb{Q}$) reached its maximum value.

For all five indicators, i.e., $\mu$, $\chi_{\gamma}$, $\mathfrak{c}_j$, $\langle\Lambda\rangle$ and $\wp_C$, in the \textit{TL}, we observed an interesting phenomenon of two \textit{first}-order phase transitions, namely, the change of phases (or, \textit{strategies}) from $\mathcal{D}\rightarrow\mathbb{Q}$ (at entanglement value $\gamma_A$) and $\mathbb{Q}\rightarrow\mathcal{D}$ (at entanglement value $\gamma_B$). This result is very similar to the ones observed in Type-\textit{I} superconductors, at a certain critical temperature and in the absence of an external field (see, Refs.~\cite{ref3, ref20}). This also showcases the fact that for QuPD, at finite entanglement $\gamma$ and \textit{zero} noise, we observe a change in the Nash equilibrium condition from \textit{All-$\mathcal{D}$} to \textit{All-$\mathbb{Q}$} and this is marked by a \textit{first}-order phase transition in all the five indicators. To conclude, this tutorial is primarily focused on mapping a \textit{one-shot} QuPD game to the $1D$-Ising chain and then numerically studying the emergence of cooperative behaviour among an infinite number of quantum players by involving five different indicators, all of which have a thermodynamic analogue.

In this tutorial, we considered three distinct strategies: Cooperate, Defect, and Quantum. However, we have also worked on playing four distinct strategies: Cooperate, Defect, Quantum and Random in a quantum game, see Ref.~\cite{ref4}. We hope to analyze the aforesaid four strategy case using agent based modelling in the future.
We also plan to further extend this formalism to \textit{repeated} QuPD games, involving unitary actions/operators like \textit{Hadamard} ($\hat{\mathbb{H}}$), etc. or other non-Unitary operators (by using modified \textit{EWL} protocol) \cite{ref8,ref25}.

\section*{Author Declarations}

\subsection{Conflict of Interest}
The authors have no conflicts to disclose.

\subsection{Data Availability Statement}
The data that supports the findings of this study are available in the article.




\begin{thebibliography}{H}

\bibitem{ref1} M.A. Nowak, S. Coakley, \emph{Evolution, Games, and God: The Principle of Cooperation} (2013)

\bibitem{ref2}M. Nowak, A. Sasaki, C. Taylor \textit{et al}, \emph{Emergence of cooperation and evolutionary stability in finite populations}, \href{https://doi.org/10.1038/nature02414}{Nature {428}, 646 (2004).}

\bibitem{ref3}C. Benjamin, A. Dash, \emph{Thermodynamic susceptibility as a measure of cooperative behaviour in social dilemmas}, \href{https://doi.org/10.1063/5.0015655}{Chaos  {30}, 093117 (2020).}

\bibitem{ref4}Jabir T. M., N. Vyas, C. Benjamin, \emph{Is the essence of a quantum game captured completely in the original classical game?}, \href{https://doi.org/10.1016/j.physa.2021.126360}{Physica A: Stat. Mech. and App. 584, 126360 (2021)}

\bibitem{ref5}A. Mukhopadhyay \textit{et. al.}, \emph{Repeated quantum game as a stochastic game: Effects of the shadow of the future and entanglement}, \href{https://doi.org/10.1016/j.physa.2024.129613}{Physica A: Stat. Mech. and App., {637}, 129613 (2024)}

\bibitem{ref7}S. Sarkar, C. Benjamin, \emph{Quantum Nash equilibrium in the thermodynamic limit}, \href{https://doi.org/10.1007/s11128-019-2237-2}{Quantum Inf. Process. {18}, 122 (2019).}

\bibitem{ref8}J. Eisert, M. Wilkens, M. Lewenstein, \emph{Quantum Games and Quantum Strategies}, \href{https://doi.org/10.1103/PhysRevLett.83.3077}{Phys. Rev. Lett. {83}, 3077 (1999).}

\bibitem{lowe} A. Lowe, Determining Quantum Correlation through Nash Equilibria in Constant-Sum Games,  \href{https://arxiv.org/abs/2410.15401}{arXiv:2410.15401}.

\bibitem{rebecca} Rebecca Erbanni, Antonios Varvitsiotis, and Dario Poletti, Entangling capabilities and unitary quantum games, \href{https://doi.org/10.1103/PhysRevA.110.022413}{Phys. Rev. A 110, 022413 (2024)}.

\bibitem{ref9}C. Benjamin, S. Sarkar, \emph{Triggers for cooperative behaviour in the thermodynamic limit: A case study in Public goods game}, \href{https://doi.org/10.1063/1.5085076}{Chaos {29}, 053131 (2019).}

\bibitem{ref10}C. Benjamin, Arjun Krishnan U.M., \emph{Nash equilibrium mapping vs. Hamiltonian dynamics vs. Darwinian evolution for some social dilemma games in the thermodynamic limit}, \href{https://doi.org/10.1140/epjb/s10051-023-00573-4}{Eur. Phys. J. B {96}, 105 (2023).}

\bibitem{ref11}C. Adami, A. Hintze, \emph{Thermodynamics of evolutionary games}, \href{https://journals.aps.org/pre/abstract/10.1103/PhysRevE.97.062136}{Phys. Rev. E {97}, 062136 (2018).}

\bibitem{ref12}R. Tah, C. Benjamin, \emph{Game susceptibility, Correlation and Payoff capacity as a measure of Cooperative behavior in the thermodynamic limit of some Social dilemmas}, \href{https://doi.org/10.48550/arXiv.2401.18065}{arXiv:2401.18065  (2024).}

\bibitem{ref19} S. Sarkar, C. Benjamin, \emph{Entanglement renders free riding redundant in the thermodynamic limit}, \href{https://doi.org/10.1016/j.physa.2019.01.085}{Physica A: Stat. Mech. and App., {521},  607 (2019).}

\bibitem{ref15} C. Benjamin, A. Dash, \emph{Switching global correlations on and off in a many-body quantum state by tuning local entanglement}, \href{https://doi.org/10.1063/5.0171825}{Chaos {33 (9)}:091104 (2023).}
\bibitem{ref17} A.P. Flitney, D. Abbott, \emph{An Introduction to Quantum Game Theory}, \href{https://doi.org/10.1142/S0219477502000981}{Fluctuation and Noise Letters, 2 (4), {R175} (2002).}

\bibitem{ref6}S. Galam, B. Walliser, \emph{Ising model versus normal form game}, \href{https://doi.org/10.1016/j.physa.2009.09.029}{Physica A: Statistical Mechanics and its Applications, {389:3} (2010).}

\bibitem{ref20} G. Jaeger, \emph{The Ehrenfest Classification of Phase Transitions: Introduction and Evolution}, \href{https://doi.org/10.1007/s004070050021}{Arch Hist Exact Sc. {53}, 51 (1998).}
\bibitem{ref5a}S. Sarkar, C. Benjamin, \emph{The emergence of cooperation in the thermodynamic limit}, \href{https://doi.org/10.1016/j.chaos.2020.109762}{Chaos {135}, 109762 (2020).}

\bibitem{ref13}P.M. Altrock, A. Traulsen, \emph{Deterministic evolutionary game dynamics in finite populations}, \href{https://journals.aps.org/pre/abstract/10.1103/PhysRevE.80.011909}{Phys. Rev. E {80}, 011909 (2009).}
\bibitem{ref14} G.T. Landi, \emph{Ising model and Landau theory, Undergraduate Statistical Mechanics, IFUSP - Physics Institute, University of São Paulo}, \href{http://www.fmt.if.usp.br/~gtlandi/lecture-notes/12---ising.pdf}{Chapter \textbf{3}, Ferromagnetism (2017).}


\bibitem{ref21} A. Sandvik, \emph{Computational Physics lecture notes, Fall 2023, Department of Physics, Boston University}, \href{https://physics.bu.edu/~py502/slides/l17.pdf}{Measuring physical observables.}

\bibitem{ref22} G. Hocky \textit{et al.}, \emph{Statistical Mechanics lecture notes, 2021, New York University}, \href{https://hockygroup.hosting.nyu.edu/exercise/ising-1d.html}{Metropolis Monte Carlo for the Ising Model.}

\bibitem{ref23} R.J. Baxter, \emph{Exactly Solved Models in Statistical Mechanics, 1982, Academic Press.}
\bibitem{jan} L. Pawela and J. Sładkowski, Quantum Prisoner's Dilemma game on hypergraph networks, \href{https://doi.org/10.1016/j.physa.2012.10.034}{Physica A 392, 910-917 (2013)}.


\bibitem{ref25} De He, T. Ye, \emph{An Improvement of Quantum Prisoners’ Dilemma Protocol of Eisert-Wilkens-Lewenstein}, \href{https://doi.org/10.1007/s10773-019-04351-w}{Int. J. Theor. Phys. {59}, 1382 (2020).}


\end{thebibliography}
\end{document}